
\input harvmac
\input epsf.tex

\def\mn{\mu\nu}
\def\sq2{\sqrt{2}}
\def\ra{\rightarrow}

\def\p{\partial}

\def\RN{Reissner-Nordstr\"om}
\def\apm{\alpha'}

\def\s42{ 2^{-{1\over 4} } }

\def\ra{\rightarrow}

\def\g{\gamma}

\def\a{\alpha}
\def\sa{r_0^2 {\rm sinh}^2\alpha }
\def\sg{r_0^2 {\rm sinh}^2\gamma }
\def\ss{r_0^2 {\rm sinh}^2\sigma }

\def\[{\left [}
\def\]{\right ]}
\def\({\left (}
\def\){\right )}

\pageno=1
\vskip 7.5truein
\centerline{\titlefont Black Holes in String Theory}
\vskip 1.0truein
\centerline{Juan Mart\'{\i}n Maldacena\foot{
e-mail: malda@physics.rutgers.edu}}
\vskip 1.1 truein

\centerline{\bf ABSTRACT} 
\bigskip\bigskip 

This thesis is devoted to trying to find a microscopic 
quantum description
of black holes. We consider black holes in  string theory 
which is 
a quantum
theory of gravity.
We find that the ``area law''  black hole entropy 
for extremal  and near-extremal  charged black holes 
arises 
from counting microscopic configurations.
We study black holes in five and four spacetime dimensions.
We  calculate the Hawking temperature and give
a physical picture of the Hawking decay process. 
\vskip 1.1 truein
\centerline{A Dissertation Presented to the Faculty 
of Princeton University}
\vskip .05truein
\centerline{in Candidacy for the Degree of Doctor of Philosophy}
\vskip .2truein
\centerline{Recommended for Acceptance by the Department of Physics}
\vskip .2truein
\centerline{June 1996}
\vskip .5truein

\vfill\eject

\noblackbox




\centerline{\bf TABLE OF CONTENTS} 
\settabs\+3...&aaa&3.3..&aaa&3.3.3..&aaaaaaaaaaaaaaaaaaaaaaaaaaaaaaaaaaa
aaaaaaaaaaaaaaa&\cr
\bigskip\bigskip\bigskip 
\+Acknowledgements&&&&&&3\cr 
\bigskip\medskip 
\+1.&Introduction&&&&&4\cr 
\medskip
\+&&1.1&Introduction&&&4\cr
\smallskip
\+&&1.2&Perturbative string theory&&&6\cr
\smallskip
\+&&1.3&String solitons and D-branes&&&10\cr
\bigskip\medskip
\+2. &Classical black hole solutions&&&&&19\cr
\medskip
\+&&2.1&Extended  $p$-brane solutions&&&20\cr
\smallskip
\+&&2.2&Oscillating strings and p-branes&&&24\cr
\smallskip
\+&&2.3&$d\le 9 $ black holes from $d=10$ strings or $p$-branes&&&27\cr
\smallskip
\+&&2.4&U-duality and quantization of the charges&&&30\cr
\smallskip
\+&&2.5&Black hole solutions in five  dimensions&&&33\cr
\smallskip
\+&&2.6&Black hole solutions in  four dimensions&&&43\cr
\bigskip\medskip
\+3.&D-brane description of black holes&&&&&&50\cr 
\medskip 
\+&&3.1&Extremal five dimensional black holes&&&50\cr 
\smallskip 
\+&&3.2&Near-extremal $5 d$  black holes and Hawking radiation&&&60\cr
\smallskip
\+&&3.3&Extremal and non-extremal four dimensional black holes&&&68\cr
\bigskip\medskip 
\+4.&Discussion&&&&&&73\cr
\bigskip\medskip
\+References&&&&&&76\cr

\vfill\eject 

\centerline{\bf ACKNOWLEDGEMENTS}
\bigskip

I am very grateful to my advisor, Curtis Callan, for teaching me
many things, sharing his ideas with me  and encouraging me.
I am also grateful to
Igor Klebanov and Andreas Ludwig for the ideas they shared with
me and the research  we did together.

\medskip

I am  also very grateful to David  Lowe, Gary Horowitz,
Andy Strominger and 
Lenny  Susskind  for very interesting discussions and collaborations
in which some of this work was done.

\medskip

I am thankful to Amanda  Peet for getting me interested in
black holes and fruitful collaboration.
I also had nice and stimulating discussions with  
Finn Larsen, Eva Silverstein, Clifford Johnson and
Jose Luis Barbon.
Thanks to all the other fourth floor dwellers,
Vijay Balasubramanian, Aki Hashimoto, Rajesh
Gopakumar, Chetan Nayak, Victor Gurarie, Marco
Moriconi, Guy Moore, Ali  Yegulalp.
I am also indebted to the people in Argentina that
introduced me to this field, Gerardo Aldaz\'abal,
Eduardo  Andr\'es, Oscar  Jofre and Carmen   N\'u\~{n}ez.

\medskip

I would also like to thank my friends outside physics,
Tom\'as, George, Laifong, Susanna, Karen, Karen,  Nicole,
Marco, Beatriz, Raul, Ramiro, Walter, Marcelo, Celso, Olga,
Jorgito, M\'onica, Marcelo, Leo, la familia Rodr\'{\i}guez, la gente de
Bariloche, 
and many others.
I also owe all the people in
the Aquinas Catholic group special
thanks.

\medskip

Finally, I thank my family for their constant encouragement.

\vfill\eject



\lref\wadia{A. Dhar, G. Mandal and S. Wadia,
{\it Absorption vs Decay of Black holes in string theory and T-symmetry},
 hep-th/9605234.}

\lref\dasmathurone{S. Das and S. Mathur, {\it
Comparing decay rates for black holes and D-branes},
 hep-th/9606185.}

\lref\dasmathurtwo{S. Das and S. Mathur,
{\it Interactions involving D-branes},
hep-th/9607149.}

\lref\kara{ R. Kallosh and A. Rajaraman, 
{\it Brane-Antibrane Democracy },
hep-th/9604193.}

\lref\schwartzmultstrings{
 J. Schwarz, Phys.Lett. {\bf B360} (1995) 13;
ERRATUM-ibid.{\bf B364} (1995) 252.}

\lref\REV{For some reviews see: M. Green, J. Schwarz, and E. Witten, 
 ``Superstring Theory,'' two volumes (Cambridge University Press,
1987)\semi
A.M. Polyakov, ``Gauge Fields and Strings,'' (Harwood, 1987) \semi
M. Kaku, ``Introduction to Superstrings,'' (Springer-Verlag, 1988) \semi
D. Lust and S. Theisen, ``Lectures on String Theory,'' (Springer-Verlag,
1989)\semi
M. Kaku, ``Strings, Conformal Fields, and Topology: An Introduction,''
 (Springer-Verlag, 1991)\semi
L. Alvarez-Gaume and M. A. Vazquez-Mozo, 
``Topics in String Theory and Quantum Gravity,'' in Gravitation and 
Quantizations, Eds. B. Julia and J. Zinn-Justin, (North Holland 1995)
hep-th/9212006. } 
\lref\hs{G. Horowitz and A. Strominger, Nucl.Phys {\bf B 360} (1991) 197.}

\lref\sv{A. Strominger and C. Vafa, {\it
 Microscopic Origin of the Bekenstein-Hawking Entropy},
 hep-th/9601029.}

\lref\cmp{C.C. Callan, J.M. Maldacena and A.W. Peet, {\it 
Extremal Black Holes As Fundamental Strings
},  hep-th/9510134,
to appear in Nucl. Phys. B.}

\lref\spn{J. Breckenridge, R. Myers, A. Peet and C. Vafa, {\it
D--branes and Spinning Black Holes}, hep-th/9602065.}

\lref\vbd{J. Breckenridge, D. Lowe, R. Myers, A.  Peet, A. Strominger and
C. Vafa, {\it 
Macroscopic and Microscopic Entropy of Near-Extremal Spinning Black Holes},
 hep-th/9603078.}

\lref\cveticang{ M. Cvetic and D. Youm, {\it 
Entropy of Non-Extreme Charged Rotating Black Holes in String Theory},
hep-th/9603147.}

\lref\tseytlin{ M. Cvetic and D. Youm, {\it Dyonic PBS saturated 
black holes of heterotic string theory on a six torus, } Phys. Rev. {
\bf D53} (1996) 584, hep-th/9507090 ;
A.A. Tseytlin, 
{\it Extreme dyonic black holes in string theory},
 hep-th/9601177. }

\lref\mirjam{ M. Cvetic and A. Tseytlin, {\it 
General class of BPS saturated dyonic black holes 
as exact superstring solutions},
 hep-th/9510097.}

\lref\tseylinother{
G. Horowitz and A. Tseytlin, Phys.Rev.{\bf D51} (1995) 2896, hep-th/9409021.}

\lref\ckmy{C.G. Callan, I.K. Klebanov, J.M. Maldacena and A. Yegulalp,
Nucl.Phys.{\bf B443} (1995) 444, hep-th/9503014.}

\lref\sentduality{ A. Sen, {\it T-Duality of p-Branes}, hep-th/9512203.}

\lref\dh{A. Dabholkar and J. Harvey, Phys. Rev. Lett. {\bf 63} (1989) 478;
A. Dabholkar, G. Gibbons, J. Harvey and F. Ruiz-Ruiz, Nucl. Phys. {\bf B340
}
(1990) 33. }

\lref\dabhargaunwald{
A. Dabholkar,
J. Gauntlett, J. Harvey and D. Waldram,
{\it Strings as Solitons \& Black Holes as Strings}, hep-th/9511053.}

\lref\waldram{D. Waldram, Phys. Rev. {\bf D47} (1993) 2528.}

\lref\callanfive{ C. Callan, J. Harvey and A. Strominger,
Nucl. Phys. {\bf B359 }
(1991) 611. }

\lref\larsen{ F. Larsen and F. Wilczek, {\it 
Internal Structure of Black Holes}, 
hep-th/9511064.}

\lref\myresperry{ F. Tangherlini, Nuovo Cimento {\bf 77} (1963) 636;
R. Myers and M. Perry, Ann. Phys. {\bf 172} (1986) 304.}

\lref\daipol{ J. Dai, R. Leigh and J. Polchinski, Mod. Phys. Lett.
{\bf A 4} (1989) 2073; P. Horava, Phys. Lett. {\bf B231}
(1989) 251.}
\lref\polchinski{J. Polchinski, Phys. Rev. Lett. 75 (1995) 4724,
hep-th/9510017.}

\lref\cghs{C. Callan, S. Giddings, J. Harvey and
A. Strominger, Phys. Rev. {\bf D45} (1992) 403.}

\lref\limitations{
J. Preskill, P. Schwarz, A. Shapere, S. Trivedi and F. Wilczek,
Mod. Phys. Lett.{\bf A 6} (1991) 2353; C. Holzhey and F. Wilczek,
Nucl. Phys. {\bf B380} (1992) 447, hep-th/9202014;
P. Kraus and F. Wilczek,
Nucl. Phys. {\bf B433} (1995) 403. }

\lref\andyunitarity{ A. Strominger, {\it
Unitary Rules for Black Hole Evaporation}, hep-th/9410187; J. Polchinski and
A. Strominger, Phys.Rev. {\bf D50} (1994) 7403,  hep-th/9407008.
}

\lref\schwartzmultstrings{
 J. Schwarz, Phys.Lett. {\bf B360} (1995) 13;
ERRATUM-ibid.{\bf B364} (1995) 252.}

\lref\hull{
C. Hull and P. Townsend, Nucl. Phys. {\bf B438} (1995) 109,
hep-th/9410167.}

\lref\horavad{ P. Horava, Phys. Lett. {\bf B 289} (1992) 293.}

\lref\schwarzhair{ J. Schwarz, Phys. Lett. {\bf 272B} (1991) 239.}

\lref\susskindu{ L. Susskind and J. Uglum, Phys. Rev. {\bf D 50}
 (1994) 2700.}

\lref\wittenbound{E. Witten, Nucl. Phys. {\bf B460 }(1996) 335,
 hep-th/9510135.}

\lref\vafacount{ C. Vafa, {\it
Gas of D-Branes and Hagedorn Density of BPS States}, 
hep-th/9511088; C. Vafa, {\it
Instantons on D-branes}, hep-th/9512078.}

\lref\wittensmall{ E. Witten, Nucl. Phys. {\bf 460 }(1996) 541,
hep-th/9511030.}

\lref\complementarity{ G. 't Hooft, Nucl.Phys {\bf B 335} (1990) 138;
L. Susskind, L. Thorlacius and J. Uglum, {\bf D 48} (1993) 3743;
Y. Kiem, H. Verlinde and E. Verlinde Phys.Rev.{\bf D52},
 (1995) 7053 hep-th/9502074; D. Lowe, J. Polchinski, L. Susskind,
L. Thorlacius and J. Uglum, Phys.Rev. {\bf D52} (1995) 6997,
 hep-th/9506138; V.  Balasubramanian and H. Verlinde,
hep-th/9512148. }

\lref\hawrad{ S.W. Hawking, Comm. Math. Phys. {\bf 43} (1975) 199.}

\lref\membrane{ K. Thorne, R. Price and D. MacDonald,`` {\it
Black Holes: The Membrane Paradigm }'', Yale University Press, 1986;
L. Susskind, L. Thorlacius, J. Uglum
Phys. Rev. {\bf D48} (1993) 3743, hep-th/9306069.
 }

\lref\hawkingentropy{ J. Beckenstein, Phys. Rev. {\bf D7} (1973) 2333;
Phys. Rev. {\bf D 9} (1974) 3292; S. W. Hawking, Phys. Rev. {\bf D13 }
(1976) 191.}

\lref\hawkinguni{ S. Hawking, Phys. Rev. {\bf D 14} (1976) 2460;
For a recent review see S. Giddings,
 {\it The Black Hole Information Paradox},
hep-th/9508151 and references therein.
}

\lref\hologram{ L. Susskind, J. Math. Phys. {\bf 36 } (1995) 6377.}

\lref\lenyspeculations{ L. Susskind, RU-93-44, {\it
Some Speculations about Black Hole Entropy in String Theory},
  hep-th/9309145.}

\lref\thkl{ I. Klebanov and L. Thorlacius,  Phys. Lett. {\bf B371}
(1996) 51,
hep-th/9510200.}

\lref\ghkm{ S. Gubser, A. Hashimoto, I. Klebanov and J. Maldacena,
 {\it Gravitational lensing by p-brane}, hep-th/9601057.}

\lref\hk{  A. Hashimoto and  I. Klebanov, {\it
Decay of excited D-branes}, hep-th/9604065.}

\lref\vafaintersecting{ M. Bershadsky, C. Vafa and V. Sadov,
{\it D-Strings on D-Manifolds},
hep-th/9510225; A. Sen, {\it U-duality and Intersecting D-branes},
hep-th/9511026.}

\lref\dgl{M. Douglas, {\it Branes within Branes}, hep-th/9512077.}

\lref\dasmathur{  S.R. Das and  S.D. Mathur, {\it 
Excitations of D-strings, Entropy and Duality},
hep-th/9601152.}

\lref\wittenpol{ J. Polchinski and E. Witten, Nucl. Phys. 
{\bf B460} (1996) 506, hep-th/9510169.}

\lref\dimensionalreduction{
J. Maharana and J. Schwarz, Nucl.Phys. {\bf B390} (1993) 3;
A. Sen, Nucl. Phys. {\bf D404} (1993) 109.}

\lref\cvyo{ M. Cvetic and D. Youm, {\it 
 BPS Saturated Dyonic Black Holes of $N=8$ Supergravity Vacua},
 hep-th/9510098.}

\lref\topological{ S. Ferrara, R. Kallosh and A. Strominger, Phys.
Rev. {\bf D 52} (1995) 5412, hep-th/9508072; M. Cvetic and D. Youm,
hep-th/9507090; G. Gibbons and P. Townsend, Phys. Rev. Lett. {\bf 71}
(1993) 3754.}

\lref\polchinskinotes{ J. Polchinski, S. Chaudhuri and C. Johnson,
{\it Notes on D-Branes}, hep-th/9602052.}

\lref\hstro{ G. Horowitz and A. Strominger, {\it
Counting States of Near-Extremal Black Holes},
 hep-th/9602051.}

\lref\hor{G.T. Horowitz and A.A. Tseytlin, Phys. Rev. D51 (1995)
  2896, hep-th/9409021.}

\lref\sm{J. Maharana and J. Schwarz, Nucl. Phys. B 396 (1993) 3,
  hep-th/9207016.}

\lref\sentendfourd{ A. Sen, Int. J. Mod. Phys. A9 (1994) 3707,
  hep-th/9402002.}


\lref\cvetd{M. Cvetic and D. Youm, {\it 
BPS Saturated and Non-Extreme States in Abelian Kaluza-Klein Theory and
    Effective $N=4$ Supersymmetric String Vacua},
hep-th/9508058; {\it General Static Spherically Symmetric 
Black Holes of Heterotic String on a Six Torus},
 hep-th/9512127.}

\lref\vespachati{ T. Vachaspati and
 D. Garfinkle, Phys. Rev. D42 (1990)1960;
D. Garfinkle, Phys. Rev. D46 (1992) 4286.
}

\lref\hortsesing{ G.T. Horowitz and A.A. Tseytlin, Phys. Rev. D50
  (1994) 5204, hep-th/9406067. }

\lref\hortseprl{G.T. Horowitz and A.A. Tseytlin, Phys. Rev. Lett.
  73 (1994) 3351, hep-th/9408040.}

\lref\ber{ K. Behrndt, Nucl.Phys.B455:188-210,1995, hep-th 9506106. }

\lref\kallosh{ R. Kallosh, D. Kastor, T. Ort{\'\i}n and T. Torma,
  Phys. Rev. D50 (1994) 6374, hep-th/9406059;
  E. Bergshoeff, R. Kallosh and T. Ort{\'\i}n, Phys. Rev. D50
  (1994) 5188, hep-th/9406009;
R. Kallosh and A. Linde, SU-ITP-95-14,
 hep-th/9507022.}

\lref\senbh{A. Sen, Nucl. Phys. B440 (1995) 421, hep-th/9411187.}

\lref\peet{A. Peet, {\it Entropy and supersymmetry 
of $D$ dimensional extremal electric black holes versus
    string states},  (to appear in Nucl. Phys. B),
  hep-th/9506200.}

\lref\GSW{M. Green, J. Schwarz, and 
E. Witten, ``Superstring theory,'' vols. 
I and II (Cambridge University Press, 1987).}  

\lref\jpup{J. Polchinski, private communication.}

\lref\rk{S. Ferrara and R. Kallosh,{\it
 Supersymmetry and Attractors}, hep-th/9602136.}

\lref\ms{J. Maldacena and A. Strominger, {\it
Statistical Entropy of Four-Dimensional Extremal Black Holes},
 hep-th/9603060.}

\lref\cliffordfd{ C. Johnson, R. Khuri, R. Myers, 
{\it  Entropy of 4D Extremal Black Holes}, hep-th/9603061.}

\lref\koll{R. Kallosh and B. Koll, {\it
 E(7) Symmetric Area of the Black Hole Horizon},
 hep-th/9602014; R. Dijkgraaf,
E. Verlinde and H. Verlinde, {\it 
BPS Spectrum of the Five-Brane and Black Hole Entropy},
  hep-th/9603126;  R. Dijkgraaf,
E. Verlinde and H. Verlinde, {\it BPS Quantization of the Five-Brane},
hep-th/9604055. }


\lref\cama{ C. Callan and J. Maldacena, {\it 
D-brane Approach to Black Hole Quantum Mechanics}, hep-th/9602043.}

\lref\hms{ G. Horowitz, J. Maldacena and A. Strominger, {\it
Nonextremal Black Hole Microstates and U-duality },
 hep-th/9603109.}

\lref\hlm{ G. Horowitz, D. Lowe and  J. Maldacena, 
{\it Statistical Entropy of Nonextremal
 Four-Dimensional Black Holes and U-Duality}, 
hep-th/9603195.}

\lref\sm{ J. Maldacena and L. Susskind, {\it  D-branes and Fat Black Holes},
 hep-th/9604042.}

\lref\asop{A. Strominger, {\it Open p-branes}, hep-th/9512059.}

\lref\twn{P. Townsend, {\it D-branes from M-branes}, hep-th/9512062.}

\lref\harstro{J.A. Harvey and A. Strominger, Nucl.
Phys. {\bf B449} (1995) 535.}

\lref\penrose{ R. Penrose, Phys. Rev. Lett. {\bf 14} (1965) 57.}

\lref\censor{ R. Penrose, ``Singularities and Time Asymmetry'' and
R. Geroch and G. Horowitz ``Global Structure of Spacetimes'', both 
in {\it General Relativity, an Einstein Centenary Survey}, ed.
S. Hawking and W. Israel (Cambridge University Press)(1979) }

\lref\genrel{ Some books on General Relativity are:
C. Misner, K. Thorne and J. Wheeler, {\it Gravitation } (1973)
Feeman and Co. USA; R. Wald, {\it General Relativity}, The University
of Chicago Press, 1984; B. Schutz, {\it A first course in 
General Relativity }, Cambridge University Press (1990);
S. Hawking and G. Ellis, {\it The Large Scale Structure of 
Space-time}, Cambridge University Press, 1973.}

\lref\sugra{E. Cremmer and B. Julia, Nucl. Phys. {\bf B159}
(1979) 141; B. De Wit and H. Nicolai, Nucl. Phys. {\bf B208}
(1982) 323.}

\lref\senpbs{ A. Sen, Mod. Phys. Lett. {\bf A10} (1995),2081,
hep-th/9504147.}

\lref\buscher{ T. Buscher, Phys. Lett. {\bf 194B}(1987) 59,
Phys. Lett. {\bf 201 B} (1988) 466.}

\lref\twoasugra{ I. Campbell and P. West, Nucl. Phys. {\bf B243 }
(1984) 112; F. Giani and M. Pernici, Phys. Rev. {\bf D30}(1984)
325; M. Huq and M. Namazie, Class. Quant. Grav. {\bf 2} (1985) 293.}

\lref\gssugra{
M. Green and J. Schwarz, Phys. Lett. {\bf B}(1982) 444.}

\lref\schslr{ J. Schwarz, Nucl. Phys. {\bf B226}(1983) 269.}

\lref\teitel{ R. Nepomechie, Phys. Rev. {\bf D31} (1985) 1921;
C. Teitelboim, Phys. Lett. {\bf B176} (1986) 69.}

\lref\shenker{ S. Shenker, in ``{\it Random Surfaces and 
Quantum Gravity}'', eds. O. Alvarez, E. Marinari and P. Windey
(Plenum, 1991).}

\lref\cakl{ C. Callan and I. Klebanov, {\it
D-Brane Boundary State Dynamics},
 hep-th/9511173.}

\lref\gamy{ M.  Garousi and R. Myers,{\it
Superstring Scattering from D-Branes}, hep-th/9603194.}

\lref\senbhstr{A. Sen, Mod. Phys. Lett. {\bf A10} (1995) 2081,
hep-th/9504147.}

\lref\gkp{S. Gubser, I. Klebanov and A. Peet, {\it
Entropy and Temperature of Black 3-Branes},
 hep-th/9602135;
A. Strominger, unpublished notes.}

\lref\klts{I.Klebanov and A. Tseytlin, {\it 
 Entropy of Near-Extremal Black p-branes}, hep-th/9604089.}

\lref\luma{J. Maldacena and A. Ludwig, {\it
Majorana Fermions, Exact Mapping between Quantum Impurity Fixed Points with
    four bulk Fermion species, and Solution of the ``Unitarity Puzzle'' },
cond-mat/9502109;
 C. Callan, I. Klebanov, A. Ludwig, J. Maldacena,
 Nucl. Phys. B422
(1994) 417, hep-th/9402113.}

 \magnification=1200\baselineskip=14pt plus 2pt minus 1pt
\hoffset=.6truein\hsbody=5.8truein\hsize=\hsbody 
\hstitle=6.5truein
\voffset=.2truein
\vsize = 8truein


\newsec{\bf   INTRODUCTION}
\vskip .2truein

\subsec{Introduction}

It has been a long-standing challege for theoretical
physics to construct a theory of quantum gravity.
String theory is the leading candidate for a quantum
theory of gravity. 
General Relativity has the
seeds of its own destruction in it, since
smooth initial data can evolve into  singular field
configurations \penrose . Classically this is not a problem
if the singularities are hidden behind event 
horizons \censor , since this means that 
nothing can come out from the region containing the  singularity.
However, Hawking showed, under very general assumptions,
that quantum mechanics implies that  black holes
 emit particles \hawrad . 
In his approximation this radiation
is exactly thermal and contains no information 
about the state of the black hole. This leads
to the problem
of ``information loss'', since particles
can fall in carrying information but what comes
out is featureless thermal radiation \hawkinguni .
 Hawking
 argued that
this would lead to non-unitary evolution, so that
one of the basic principles of quantum mechanics
would have to be modified.

Black holes are thermal systems that obey the laws
of thermodynamics \hawkingentropy . In fact, they 
have an entropy
proportional to the area of the event horizon.
The area of the horizon is just a property of
the classical solution, it always increases
in classical processes like the collision of
two black holes. 
In most physical systems the thermodynamic entropy
has a statistical interpretation in terms of counting
microscopic configurations with the
same macroscopic properties, and in most cases
this counting  requires an understanding of the
quantum degrees of freedom of the system.
For black holes this has been a long-standing
puzzle: what are the degrees of freedom that
the Hawking-Beckenstein entropy is counting?

String theory, being a theory of quantum gravity \REV ,
should be able to describe black holes. 
Difficulties were very soon encountered because 
black holes involve strong coupling and therefore
one will have to go beyond simple perturbative
string theory to describe them. 
Recently there has been remarkable progress in
understanding some 
 string solitons called ``D-branes'' \daipol ,\polchinski 
,\polchinskinotes .
They account for some non-perturbative
effects in string theory and they have a very simple
description.

Charged black holes in General Relativity are characterized
by their mass $M$ and charge $Q$. The condition that the
singularity is hidden behind a horizon implies that
$M \ge Q$.
The case of $ M=Q$ is called extremal \genrel .
These black holes have smooth geometries at the horizon
and a free-falling observer would not feel anything as he
falls through the horizon. The horizon area, and thus the entropy,
are nonzero, both for the extremal and non-extremal cases.
The Hawking temperature vanishes for the extremal case and
it increases as we increase the mass  moving
away from extremality (keeping $Q$ fixed).
  For very large mass it decreases again. 

We will be considering black holes in a theory,
called N=8 supergravity \sugra ,  that is
not precisely  usual General Relativity but that is
very similar for the kind of problems we are interested in.
 The  difference with 
General Relativity is that it  contains many extra massless
fields:
U(1) gauge fields, scalar fields and various  
 fermionic  fields. Despite this different 
field content, there is  a charged  black hole solution that
is  exactly like the one in General Relativity: 
the  metric is exactly the same, there is only one
gauge field excited (which is a particular combination
of the original ones) and the rest of the fields, including 
the scalars, are all zero.
This implies, as in General Relativity, that the geometry
at the horizon is smooth. 
$N=8$ supergravity in four dimensions is  the
low energy limit of type II string theory compactified
on a six-torus $T^6$. String theory contains 
``D-brane'' solitons that are extended membranes of
various possible spatial 
dimensions \polchinski , \polchinskinotes .
 When these 
extended branes are wrapped around the compact 
directions they appear  to the four-dimensional 
observer as localized objects, as charged particles.
There is a symmetry, called U-duality, that 
interchanges all these objects \hull  . 
Superimposing many 
of these objects of different dimensions
we obtain a string soliton that has many of
the properties of a black hole \sv ,\cama ,\spn ,\vbd ,\hms ,\hstro ,\ms 
,\cliffordfd ,\hlm .
There is a large degeneracy which gives
a statistical interpretation to the 
thermodynamic entropy. 
One great virtue of considering this supergravity theory
is that the extremal black holes become supersymmetric 
configurations so that certain quantities can be calculated
at weak coupling and are then valid for all values of the
coupling. This has been the key to providing
a precise calculation of extremal 
black hole entropy. 
The entropy calculated using the ``D-brane'' method
agrees precisely, including the {\it numerical } coefficient
with the classical Hawking-Beckenstein ``area law'' \sv ,\spn ,\ms
,\cliffordfd .
The near-extremal black holes can also be considered
from this point of view and they correspond to excited
states of the solitons. These excited states result from
 attaching open strings to the 
D-branes \daipol  \polchinski . 
Hawking radiation is described by open strings
colliding and forming a closed string that leaves the
soliton \cama . Doing an average over the 
initial state of the black hole we get thermal
Hawking radiation with the correct value for
the temperature and the 
radiation rate is proportional to the area of the black hole \cama . 
These  near-extremal calculations stand on a more shaky ground
since one does not have supersymmetry to protect 
the calculations from strong coupling problems.
The successful calculation of the
entropy gives   evidence in favor of the
proposed physical picture.
Unfortunately, these uncontrolled
approximations for the near-extremal case will
prevent us from saying anything about 
 the information loss problem, but deeper
analysis of this model might lead to an answer
to this elusive problem.

In the rest of this chapter we 
review  some general facts
about string theory and we  introduce the 
string solitons called ``D-branes''.
In chapter 2  we  
describe the classical black hole supergravity solutions.
Using some string theory information about 
the quantization and nature of the various
charges, we rewrite
the entropy formulas in a very
suggestive form  in terms of basic constituents.
In chapter 3  we will show how to derive these
entropy formulas for the extremal case and  then
consider near-extremal black holes,
suggesting a physical picture for black holes
in terms of D-branes.
We conclude with a discussion on the results.


\subsec{Perturbative string theory.}

String theory is a quantum theory of interacting relativistic
strings. Much of what we can presently do involves   treating  
this interaction in perturbation theory \REV . 
But before we say anything about interactions let us review some
properties of free string theory. 
We will  be considering the theory of closed oriented strings.
The free string action is
\eqn\actionstr{
S = { 1 \over 4 \pi \alpha' } \int d^2\sigma  \[
 \p_\alpha X^\mu  \p^\alpha  X_\mu
+ \bar {\psi}^\mu \not \!\!  \p \psi_\mu  \]
}
where $ T = { 1 \over 2 \pi \alpha' } $ is the string tension.
We also have to impose the additional constraint that the two
dimensional supercurrent and stress tensor associated to \actionstr\
vanish \GSW . In this fashion 
the  bosonic part of the action, which involves the ten 
spacetime coordinates $X^\mu$, 
is just proportional to the area of the worldsheet embedded in 
ten dimensional space. The 
string contains  fermionic degrees of freedom living
on the worldsheet $\psi^\mu$.
Depending on the   boundary conditions
 of the fermions when they 
go around the loop there are four sectors which correspond
to whether the left and right moving fermions are periodic
or anti periodic as we go around the loop. 
The spacetime 
bosons come from the sectors where the boundary conditions
for the fermions are the same both for left and right moving
strings. They are the (NS,NS) and the (R,R) sectors, NS stands
for Neveu-Schwarz and R for Ramond. 
The  (NS,NS)  sector contains massless fields corresponding
to a graviton, a two form or antisymmetric tensor $B_{\mu\nu}$ and
a scalar, the dilaton $\phi$. The (R,R) sector contains antisymmetric
tensor fields of various number of indices, i.e.  $p+1$ 
 forms $A_{p+1}$.

Spacetime symmetries correspond to symmetries in the worldsheet 
conformal field theory. In some cases the symmetry comes
from a primary field conserved current.
This is the case for translations and for supersymmetry transformations.
The translations are associated to the primary fields $\p X^\mu $ and
the supersymmetries to the fermion vertex operators at zero momentum
$V_\alpha (z)$.
 This is the operator that,
in CFT,  switches between Ramond and NS sectors, as a spacetime
supersymmetry should do.
The left and right moving spinors on the world sheet can
have the same or opposite ten dimensional chiralities,
giving the IIB or  IIA theory respectively.

It will be interesting to consider strings on compact spaces.
We will concentrate on the simplest compactification which is
called toroidal and is obtained by identifying  one the 
coordinates as $ X^9 \sim X^9 + 2 \pi R $ \GSW . 
In this case the momentum $P^9$ becomes quantized in units
of $1/R$, $P^9 = n/R $. The string can also wind 
along this compact direction so that when we go around
the string  the coordinate has to satisfy the
condition $X^9 \rightarrow X^9 + 2 \pi R m$.
The two integers $(n,m)$ are the momentum and winding numbers of the
string. The Virasoro constraints are
\eqn\virasoro{\eqalign{
E^2 = & \vec P^2 + \left( { n \over R } - { m R \over \alpha' } 
\right)^2 + { 4 \over \alpha' } N_L ~,\cr
E^2 = & \vec P^2 + \left( { n \over R } + { m R \over \alpha' }
\right)^2 + { 4 \over \alpha' } N_R~,
}}
where $\vec P$ is the momentum in the directions $1,..,8$ and
$N_{L,R}$ are the total net oscillator level of the 
string\foot{
We are calculating $N_{L,R}$   it in the light front gauge, so there is
no shift in $N_{L,R}$ .}.
Combining both equations in \virasoro\
we get the level matching condition
\eqn\matching{
 P_{9 R }^2 - P_{9 L}^2 = 4 n m = 4 (N_L - N_R)~,
}
where $P_{9 L,R}= { n \over R } \mp { m R \over \alpha' }$ 
are the left and right moving momenta in the
direction $9$.
Momentum and winding are conserved and 
 appear as charges in the extended dimensions
$0,..,8$. In fact, from the $1+8$ dimensional point of view 
 they are  central charges because
they appear in the supersymmetry algebra.
The reason they appear in the supersymmetry algebra is that
they appear in the ten dimensional algebra as the left and
right moving momenta.
\eqn\susyten{
\{ Q_\alpha^L, Q_\beta^L \} = \Gamma_{\alpha \beta}^\mu 
P_{\mu L}~,~~~~~~~~~~~
\{ Q_\alpha^R, Q_\beta^R \} = \Gamma_{\alpha \beta}^\mu P_{\mu R}~.
}
The supersymmetry algebra implies that $P^0 \ge |\vec P_L |,~
P^0 \ge  
| \vec P_R | $. These are  the so called Bogomolny bounds.
If any of these   bounds is saturated we can see from \susyten\ that
some supersymmetries annihilate the state. 
If both bounds are saturated, then 
we have pure momentum or pure winding, 
$N_L = N_R =0$, and  half of the supersymmetries are broken.
If only one bound, let us  say the one involving $P_R$, is saturated, 
then $N_R =0$, $N_L$ is given by \matching\  and 
only 1/4 of the supersymmetries are left unbroken. 

We can see from \virasoro\ that the spectrum is left invariant
under the change of  $  R  \rightarrow  \alpha'/R  $.
This turns out to be a symmetry of the whole string theory,
 also of the interactions,
and  we expect that it will be valid even non-perturbatively.
This  very  important symmetry of  string theory is 
called T-duality. In fact in order for it to be a symmetry of
the interactions also we need to change the 
coupling constant together with the radius as \buscher 
\eqn\tduality{
R \rightarrow R' = { \alpha' \over R } ~,~~~~~~~~~~~~~~~~
g \rightarrow g' = { g \sqrt{\alpha'} \over R }~.
}
The change in the coupling constant is such that the 
$d$ dimensional Newton constant stays invariant.
\vskip 1cm
\vbox{
{\centerline{\epsfxsize=4in \epsfbox{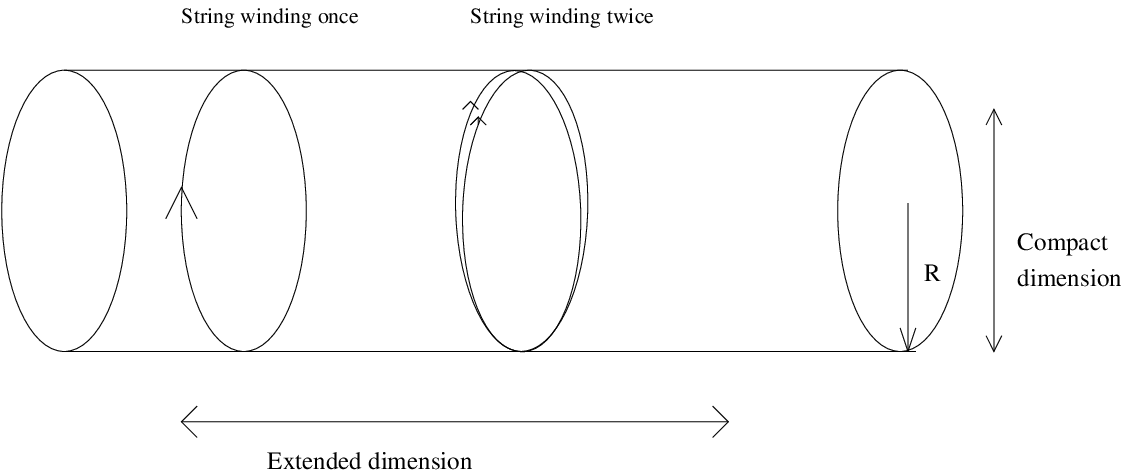}}}
{\centerline{\tenrm FIGURE 1:
Strings winding around a compact direction. }}
}
\vskip .5cm

Since string theory contains some massless particles, separated
by a large mass gap $ 1/\sqrt{\alpha'} $ from the massive
states of the string, it is natural to study 
the effective low energy  action describing the interacting
string theory.  It has to have the symmetries of the theory:
N=2 local supersymmetry in ten dimensions.
To lowest order in string perturbation theory and long
distances (we keep only second derivative terms) 
 this Lagrangian is that of type II 
ten dimensional supergravity. It is called  type II because we have two
supersymmetries. Depending on the relative chirality of
the supersymmetry generators we have the type IIA or IIB theories,
which are the limits the IIA or IIB  string theories.
Let us start with the 
ten dimensional type IIA supergravity action \twoasugra . 
This theory contains the fields coming from the 
 (NS,NS) sector  which are
the metric $G_{\mu\nu}$, a two form
$B_{\mu\nu}$  and a scalar
$\phi $  called the dilaton. The fields coming from the
(R,R) sector are 
 a one form $A_\mu$ and a  three form $ C_{\mu\nu\rho} $.
It also has
the supersymmetic fermionic partners of all
these fields.
The bosonic part of the action is
\eqn\actiontwoa{
\eqalign{{S} &={1 \over 16 \pi G_N^{10}  }
{\int} {d^{10}}{x}{\sqrt{-G}}\biggl[ {e^{-2\phi}}({R+4}
({\nabla}{\phi})^2
-{1\over3}{H^2})-\apm {G^2}-{\apm \over12}{F^{\prime 2}}\cr
&~~~~~~\qquad \qquad  \qquad \qquad
-{\apm \over 288}\epsilon^{\mu_1...\mu_{10}}
F_{\mu_1\mu_2\mu_3\mu_4}F_{\mu_5\mu_6\mu_7\mu_8}
B_{\mu_9\mu_{10}}\biggr]
}}
where $G=dA,~~H=dB$,  $F=dC$ and $F^{\prime}=dC+2A\wedge H$ are
the field strengths
 associated with each of the differential forms.\foot{In
components $G_{\mu\nu}=2\partial_{[\mu}A_{\mu]},~~
H_{\mu\nu\rho}=3\partial_{[\mu}B_{\nu\rho]},~~
F^{\prime}_{\mu\nu\rho\sigma }=4\partial_{[\mu}C_{\nu\rho\sigma]}+
8A_{[\mu}H_{\nu\rho\sigma]}$.}
The supersymmetries are generated by two spinors $\epsilon_{L,R}$ of
opposite chirality.
The gravitational part of the action can be put in the
standard form ($ S \sim \int \sqrt{g} R $) by defining a new metric, called
 the
Einstein metric, by $ g_E = e^{- \phi/2} G $, where $G$ is the 
(so called) string  
metric in \actiontwoa .

The type IIB action contains the same fields coming from the 
(NS,NS) sector and it contains therefore the first three
terms in the action \actiontwoa . The (R,R) fields are
a scalar
$\chi$ (or zero form), a two form
$B'_{\mu\nu} $ and
and four form $A_{\mu\nu\rho\delta}$ 
whose field strength is self dual $F= dA = * F$.
Due to this condition
 it is not possible to write a covariant action for
the IIB theory, however the equations of motion and
the supersymmetry variations are known.
We can also truncate the theory setting $F=0$ and then
we have a covariant  action for the rest of the fields.

Type IIB supergravity \gssugra\ 
 has the interesting property that
it is S-dual  under changing $ \phi \rightarrow - \phi$ and
interchanging the two antisymmetric tensor fields $
B \leftrightarrow B' $ \schslr. In fact, the classical symmetry is
SL(2,R) once one includes shifts in the other scalar $\chi$.
In string theory an SL(2,Z) subgroup of this symmetry
is expected to survive \hull . We will later make use of
this S-duality symmetry to generate solutions and
relate them to each other. 
This S-duality transformation leaves the Einstein metric 
invariant but it changes the string metric. 
This means, in particular, that if we have a compactified theory
and the radii are measured in string metric as in 
\virasoro\ then, under an S-duality transformation they 
all change as 
\eqn\sduality{
g \rightarrow g' = { 1\over g} ~,~~~~~~~~~~~~~~
R_i \rightarrow R_i  = { R_i  \over \sqrt{g} } ~.
}
We will define throughout this thesis the ten dimensional coupling
constant $g= e^{\phi_{\infty} } $ 
to be such that it transforms as \sduality\ under S-duality.
We will see in section (2.4)  that this fixes the ten dimensional
Newton constant in \actiontwoa\ to be 
$ G^{10}_N = 8 \pi^6 g^2 \alpha'^4  $.
In compactified theories the S-duality and T-duality groups
combine to form a bigger group called U-duality \hull .

\subsec{String solitons and D-branes.}

The low energy supergravity action contains $p+1$ forms
$A_{p+1}$ 
coming from the RR sector, $p$ is even in IIA, and odd in IIB.
 There are no objects in perturbative string theory that
carry charge under these fields, all vertex operators 
involve the field strength of these forms. The objects that
would carry charge under a $p+1$ form are extended $p$ branes.
The coupling  is 
\eqn\coupling{
 \mu_p \int_{V_{p+1} } A_{p+1}
} 
which naturally generalizes the electromagnetic coupling to
an electric charge. 
In addition, if we assume that the spectrum of electric
$p$ brane charges is quantized we would expect also 
``magnetic'' $6-p$ branes that couple to the 
Dirac dual $\tilde A_{7-p}$ form defined through 
equations of the type $ d \tilde A_{7-p} = \ast  d A_{p+1} $ (the
details are slightly more complicated) \teitel . 

In fact,
 type II supergravities contain extended black $p$ brane
solutions which carry this charge \hs . 
The extremal limit of these $p$ branes saturates the
BPS bound for these charges. These solutions will be
presented in chapter 2. 

In 
string theory these solutions appear as some very 
special solitons \daipol \polchinski .
They are extended objects with $p$ spatial dimensions and
are called ``D-branes''.
Their description is very simple and it amounts
to the following definition: 
{ \it D-branes are p-dimensional extended surfaces in spacetime 
where strings can end.} A D-brane is the string theory
solution (it is described by  a CFT)
 whose low energy limit is a supergravity 
extremal $p$ brane. 
In type II theory we had only closed strings in the
vacuum. In the presence of a D-brane there are also
open strings which interact with the closed strings
by usual splitting and joining interactions \GSW . 
These D-branes have the peculiar property that their
mass (tension) goes like $ 1/g $ and in fact they would
lead to  non-perturbative
effects of  order $e^{-O(1/g)}$. Effects  of this 
 magnitude were observed in string
theory, specially in matrix models \shenker . They also carry 
RR charges with the values predicted by U-duality.

An open string has a worldsheet that is topologically a
strip. One has to specify some boundary conditions
on the boundaries of the strip, that is, at the
end of the string.
The boundary conditions describing an open string 
attached to a $p$ brane sitting at $x_{p+1} = \cdots = x_9 =0$
 are
\eqn\dirich{\eqalign{
\p_\sigma X^\mu =& 0~~~~~~~~~~{\rm for }~~~ \mu = 0,...,p~,
\cr
X^\mu =&0 ~~~~~~~~~~{\rm for }~~~ \mu = p+1,...,9~.
}}
These are Neumann boundary conditions on the directions
parallel to the brane and Dirichlet conditions on 
the directions perpendicular to the brane. 
This is the reason they are called D(irichlet)-branes.
These open strings have the characteristic
spectrum $P^2= { 4 \over \alpha' } N_{open} $, with the
momentum $P= (P^0,...,P^p)$ being parallel to the 
brane. These open strings 
 represent excitations of the branes. In general,
an excited brane corresponds to having a gas of 
these open strings on the brane. 
Of particular interest to us will be the massless bosonic
open strings, those for which $N_{open} =0 $.
The massless open strings have a vector index. If the
index lies in the directions parallel to the brane
they describe gauge fields living on the brane and if
the index is perpendicular to the 
brane they describe oscillations of the brane in the
perpendicular directions.
As an example let us take a D-string, consider it winding
once around the compact direction $\hat 9$. Note that 
S-duality interchanges this D-string with a
 fundamental string \hull \schwartzmultstrings .
  The open strings attached to the
D-string  
can have momentum in the direction $\hat9$ which is  quantized in
units of $1/R_9$. The energy of a D-brane containing 
a gas of massless open strings is
\eqn\oscild{
E = { R_9 \over \alpha' g } + \sum_i \epsilon_i = 
E_0 + {N_L + N_R  \over R_9 }~.
}
For each momentum $n$ we have eight bosonic and eight fermionic
modes. There can be a number $N_n$ of strings with momentum $n$ 
and 
$$N_R = \sum_{n > 0} n N_n~,~~~~~~~~~~~~~~
N_L = \sum_{n <0 } n N_n~. $$
We see that the 
 spectrum is the same as the one we would obtain for 
 a superstring winding
around the 9$^{th}$ direction 
with tension $T_D = { 1 \over 2 \pi g \alpha' } $ 
if we expand \virasoro\ in powers of $R_9$.
 Note that the statistics  and number of
excitations corresponds precisely with that of the
fundamental string. 
In this  fashion we can see that the massless 
open strings describe oscillations of D-branes.
Actually, for $p >1$  not only oscillations 
but also fluctuations
in the world-brane gauge fields.
\vskip 1cm
\vbox{
{\centerline{\epsfxsize=4in \epsfbox{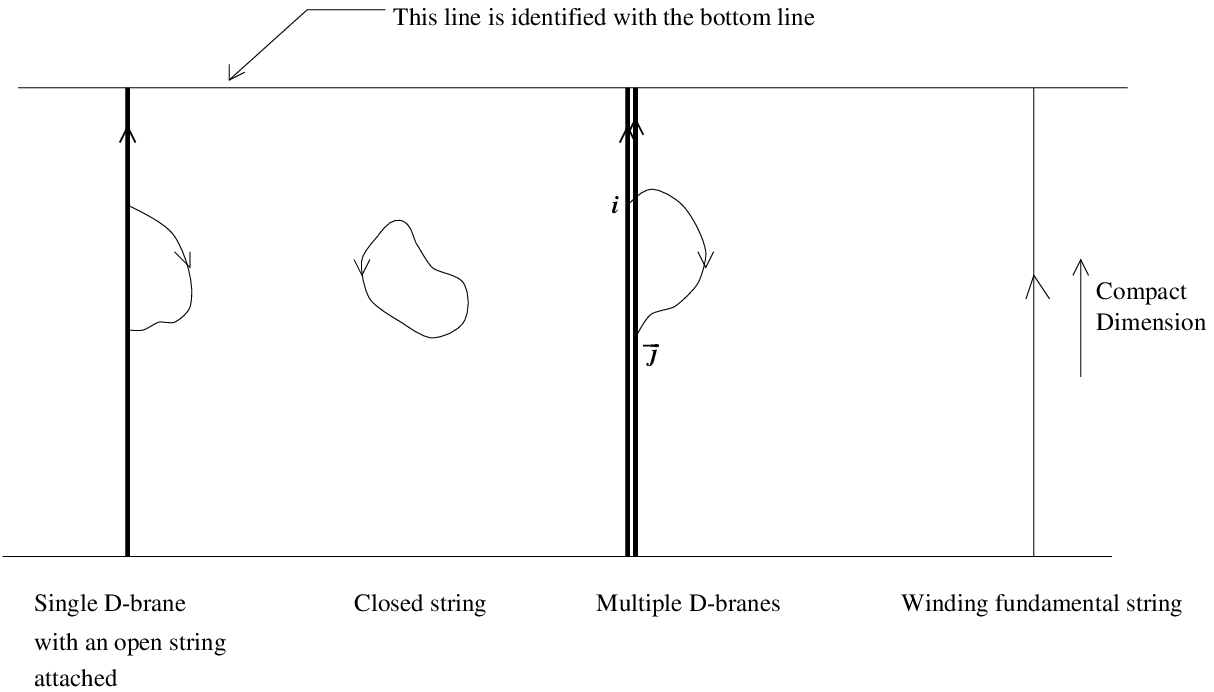}}}
{\centerline{\tenrm FIGURE 2:
D-branes winding around a compact direction with 
open strings attached.}}
{\centerline{ Only closed strings exist between widely 
separated D-branes. Open strings 
}}
{\centerline{
 carry U(N) Chan Paton factors when we have
several D-branes.
}}
}
\vskip .5cm

If one considers many D-branes of the same
type sitting on top of each other, the
open strings  carry  Chan-Paton
indices $(i,\bar{j})$ specifying the starting and
ending point of the string \polchinskinotes \wittenbound . 
The interactions of these massless open strings 
can be described by a U(N) Yang-Mills action. 
Since T-duality transformations change the dimensionality
of D-branes
the simplest way to obtain this action  is to do a T-duality 
transformation into  9-branes filling the
space and  we have an N=1 ten dimensional YM Lagrangian
\eqn\tendym{
S = { 1 \over 4 g } \int d^{10} x  Tr[ F_{\mu \nu} F^{\mu \nu} ] + 
{\rm fermions }~.
}
This Lagrangian  describes the low energy limit
of open string amplitudes.
If we perform T-duality transformations the 
amplitudes will not change. The massless 
 vertex operators 
change a little, the vertex operators for  coordinates with
Neumann boundary conditions  involve the
derivative along the boundary $ \p_t X $ while the ones for
coordinates with  Dirichlet conditions  
involve the normal derivative $\p_n X $. T-duality 
interchanges the normal and tangential derivative.
Another difference is that the momentum perpendicular to the
branes vanishes. 
Otherwise the amplitudes are exactly the same. 
So we conclude that the low energy action
describing the interaction of the massless modes on
a D-brane is just the dimensional reduction of \tendym\ to $p+1$
dimensions.
So we replace $dx^{10} \rightarrow d^{p+1} x $, the 
fields $A_\alpha$, $\alpha=0,..,p$, are gauge fields
on the D-brane and $A_I$, $I=p+1,..,10$, are related to 
the motion  of the D-brane in the transverse dimensions. 
Separating the branes corresponds to breaking the symmetry 
down to U(1)$^N$ by giving an expectation value to the fields
$A_I$, $I=p+1,..,10$. These expectation values have to be
commuting diagonal matrices (up to gauge transformations), the
elements on the diagonal represent the position of the branes
\daipol \wittenbound .
In the case of a fundamental string we can have many different
configurations depending on how the string is wound in 
the compact direction. We could have a single string
wound $N$ times or $N$ strings each winding only once. 
For D-branes  we have a similar situation. Different windings
correspond to different boundary conditions along
the compact direction. 
The physics will be different 
depending on how they are wound. 
For example, if we have a single  D-string winding 
$Q$ times all the fields will satisfy the boundary 
condition $A_\mu( x_9+ 2 \pi R_9) = U A_\mu(x_9) U^{-1} $ where
$U$ is the transformation that cyclically permutes the 
 Chan-Paton indices $i \rightarrow i+1 $. 
Now we are interested in finding states of the system 
corresponding to oscillating D-strings.
Naively we might think that $Q$ D-strings have a set of
$Q^2$ independent massless excitations, corresponding to the
different components of the gauge field. However we should be 
more careful because there are interactions, so if we 
consider, for example, a configuration with waves along
the diagonal  $Q$ directions
corresponding to separating the D-strings, then the other components
of the gauge field become massive. 
In other words, in the worldbrane  gauge theory there
are 8 scalars in the adjoint $A_I$ and there is a 
potential for these  scalars, coming from the commutator
terms in the YM action, $ V= \sum_{IJ} Tr [A_I,A_J]^2 $.
In order to see this more explicitly let us take diagonal 
matrices
\eqn\aoscill{
A_I = \pmatrix{ f_I^1(u,v) & & & \cr
&.& &\cr & & .& \cr & & & f_I^Q(u,v) }~,
}
where $v,u = x^9 \pm x^0$. If we insert this ansatz in the
equations of motion we find that $f_I^m$ obey the massless
wave equation. Now  consider, on this background, a
small off diagonal component $
( \delta A_I)_{mn} \not = 0   $ ,where $m \not = n $ are
some fixed indices, and all other components of $\delta A $ are 
zero. The equation of motion will be of the form
\eqn\offdia{
4 \p_u \p_v  ( \delta A_I)_{mn}  - 
( f_J^n - f_J^m )^2  ( \delta A_I)_{mn} =0~.
}
We see that the oscillating 
background acts like a mass term for this
off diagonal component. The effect of this mass term is more 
clear if we consider purely left moving excitations. Then 
we see that the maximum number of independent oscillations  is
 $8 Q$, corresponding  to diagonal matrices $A_I$, since
the equation \offdia\ cannot be solved with purely left 
moving 
 excitations if $f_J^n$ are arbitrary. 
In the case that the $f_J^n$ contain both left and right 
moving waves it is reasonable to assume that for generic
$f's$ we are not going to have any resonances and that 
off diagonal excitations will be effectively massive. 

In the case the D-string is multiply wound 
these  diagonal elements $f^n_I$  are cyclically permuted in going
around the compact direction $f^n_I(x_9 + 2 \pi R_9) = f^{n+1}_I(x_9 )$
 so that we could think
that the momentum is quantized in units of 
$1/QR_9$. This correctly  reproduces the 
energy  levels of a multiply wound string \dasmathur\ 
\eqn\energmult{
E = { R_9 Q \over g \alpha' } + { N'_L + N'_R \over Q R_9 }~.
}
The total physical momentum still has to be quantized in units
of $1/R$ so   
$P = (N'_L-N'_R )/QR_9 = N/R_9$. This is the condition 
analogous to \matching . Here we have assumed that 
$R_9$ is very big so that we can neglect interactions and
massive open strings. 

The states with  $N_{R} =0$ are BPS and supersymmetry
ensures that 
\energmult\ is precisely 
right. 
This configuration is related by S duality to a 
 fundamental string of winding number  $Q$ 
carrying left moving oscillations.
We can see that the degeneracies are precisely the
same since we have eight bosonic and fermionic excitations
with momenta quantized in units of $1/RQ$.
 It was
crucial to obtain the reduction of the independent
degrees of freedom from $ Q^2 \rightarrow Q$. 
We will see this mechanism working again for the black hole case.

It is quite straightforward to compute the interactions 
of these open strings \cakl , the interactions of closed strings and
open strings \hk\  and the scattering of closed strings from the 
D-brane \thkl \ghkm \gamy . To lowest order in string perturbation theory
they reduce to
calculations on the disc with vertex insertions at the 
boundary associated with open strings and insertions in the
interior of the disc associated to closed string states. 
In this way we can compute the scattering of closed strings
from the D-brane and we indeed find that in the low energy limit
the stringy  amplitudes agree with those calculated purely in 
 the
supergravity $p$-brane solutions 
\ghkm \hk .
\vskip 1cm
\vbox{
{\centerline{\epsfxsize=4in \epsfbox{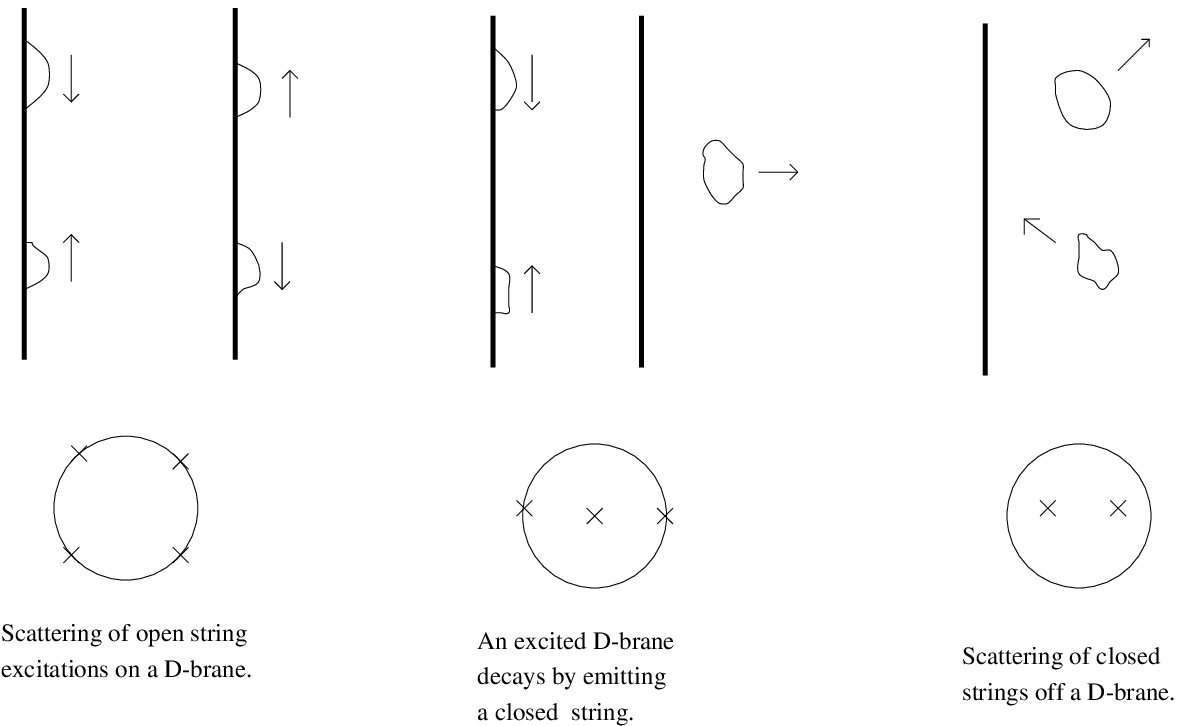}}}
{\centerline{\tenrm FIGURE 3: 
String theory diagrams appearing in various scattering
processes.}}
{\centerline{The second process is the relevant one  for Hawking radiation.
}}
}
\vskip .5cm

In the presence of a D-brane it is easy to see 
how 
supersymmetries are broken. We said before that the right
and left ten dimensional  supersymmetries are generated by the 
right and left moving spinors on the worldsheet.
The presence of a boundary in the world sheet relates the
left and right moving spinors through a boundary condition.
This is something familiar from open string theories, which
have only one supersymmetry in ten dimensions (Type I).
As argued in \ghkm \polchinskinotes\ the boundary condition 
for the spinors is
\eqn\bcspinors{ \left.
S_{R} (z) = \pm
 \Gamma^0 \cdots \Gamma^p S_L( \bar z ) \right|_{ z = \bar z }~.
}
The two choices of sign in \bcspinors\ corresponds to 
opposite D-brane orientations and therefore opposite D-brane
charges.
Note that in the type IIA theory we have $p$ even and therefore
different chiralities for the worldsheet spinors, while for
IIB theory we have odd $p$ and the same chirality for both
spinors. 
This in turn translates into the following condition
for the parameters that generate the unbroken supersymmetries
in the presence of a $p$ brane
\eqn\susydbrane{
\epsilon_R = \Gamma^0 \cdots \Gamma^p \epsilon_L ~.
}

Since the BPS  $p$-brane solution is the extremal 
limit of a black $p$-brane we would expect that  
D-branes provide a quantum description for these black
branes. This  naive expectation is not quite so
because 
the Schwarschild radius of a 
D-brane is of order $r_s^{7-p} \sim   g $ which  is much smaller than
$\sqrt{\alpha'}$ for small $g$. So the strings are typically much 
larger than the black hole radius \polchinskinotes . 
We might try to solve this by considering many D-branes, in
that case the Schwarschild radius would grow like $  r_s^{7-p} =
 Q g $. However in any process we consider there will
be open string loop corrections which will be of 
order $ g Q$, the extra factor of $Q$ comes from the
sum over the Chan Paton index. 
If we compactify the D-brane to make a black hole we 
see that  the supergravity solution already shows that
there are scalar fields that are blowing up  as we approach the
horizon, this also indicates that near the brane the strings are not
free any more and also that these black holes are very different than
the ones  we are used to in General Relativity.
Of course the size of loop corrections depends on  where the 
$Q$ D-branes are, if they are sitting on top of each other the
corrections are big but if they are separated
in space the corrections are small.
From the point of view of string theory, separating the
branes in space means giving an expectation value
to the translational zero modes of the brane, which 
means putting many open strings on the D-brane.

We will show in what follows that there are some 
properties of  black holes that are correctly described by 
D-branes. But in order to describe those black holes we need
configurations with more than one type of D-branes.

If we introduce another type  of D-brane we have even more types 
of open strings. We would like to choose $p$-brane and
$p'$-brane superpositions 
in such a  way that some supersymmetries are  still 
preserved. The additional boundary will introduce a 
new condition on the spinors of the type \susydbrane\ with
$p \rightarrow p'$. We can see that if 
$p-p' = 4,8 $ we can have a supersymmetric configuration
preserving 1/4  of the supersymmetries if the $p'$-brane
is parallel to the $p$-brane \dgl . Other configurations 
with non-parallel D-branes can be obtained from this one
by applying T-duality transformations.
The different branes need not be on top of each other and  wherever
the branes are, we have a supersymmetric configuration that
saturates the BPS bound $ M = c_p Q_p + c_{p-4} Q_{p-4} $, 
where $c$'s are some fixed coefficients.
If $p-p'=2,6$ then two conditions of the type
\susydbrane\ seem to be in conflict because they impose
chirality conditions that cannot be satisfied for
real spinors. 
Nevertheless BPS   configurations carrying
$p-2$ and $p$ brane charges   are
 predicted by U-duality, this
basically comes from the 
fact that fundamental strings can be bound to D-strings
\wittenbound . But the BPS formula for this case  \wittenbound\
has a different structure, 
$M \sim \sqrt{c'_p Q_p^2 + c'_{p-2} Q^2_{p-2} }$,
 with nonzero binding energy 
and suggests that we indeed  should  not
be able to see this configuration as two separate D-branes 
in equilibrium  at weak coupling.

If we have $Q$ coinciding D-$p$-branes, there are instanton
solutions of the U($Q$) world-brane-volume  gauge theory
with dimension $p-4$ which carry RR $p-4$ brane charge.
In fact the D-$(p-4)$-brane corresponds to the zero size 
limit of these instantons \dgl .

Intersecting D-brane
configurations with $(p,p')=(1,5) $ and $(2,6)$ will 
appear when we describe five and four dimensional black  holes.
In these cases the low energy worldbrane field theory 
describing the  interactions of the massless modes
is the dimensional reduction of 
 an $N=1$ theory in six dimensions, corresponding to
the case $(p,p') = (9,5) $ \dgl . In chapter 3 we will 
study this case in more detail.

\vfill\eject

\newsec{\bf CLASSICAL BLACK HOLE SOLUTIONS}

\vskip .2truein

In general relativity plus electromagnetism 
there are  charged black hole solutions. 
They are the most general spherically symmetric, 
stationary solutions and are characterized by the
charge $Q$ and the mass $M$. 
The cosmic censorship hypothesis \censor\  which says that 
gravitational collapse does not lead to naked
singularities 
implies that in physical situations only $M\ge Q$ black
holes will form, since the solution would 
otherwise contain a  naked singularity.
 The case $M=Q$ is called extremal, since
it has the minimum possible mass for a given charge.
This charged black holes are given by the \RN\
solution \genrel\
\eqn\renof{
ds^2 = - \Delta  dt^2 + \Delta^{-1} dr^2 + r^2 d \Omega_2^2 ~, }
$$ \Delta = \left( 1 - {r_+ \over r } \right)
 \left( 1 - {r_- \over r } \right)~.
$$
The outer horizon is at $r = r_+ $ and the mass and charge are
\eqn\massqf{
M = { 1 \over 2 G_N^4 }(r_+ + r_- )~,~~~~~~~~~~~~~~~
Q = { 1 \over G_N^4 } \sqrt{ r_+ r_- } ~.}

In this chapter we will find black hole solutions
to type II supergravity compactified down to 
$d = 4,5$ dimensions. For $d=4$ this leads to 
 $N=8$ 
supergravity. 
 The familiar solution 
\renof\ will be indeed one particular case of 
the black holes we consider. Of course,
the theory in which it is embedded is different
but the metric is the same and the gauge field
will be a particular linear combination of the
ones appearing in $N=8$ supergravity.  
These black holes can be thought of as 
extended membranes wrapping around internal 
dimensions. We will therefore start by studying
the extended brane solutions in ten dimensions.
In the following section we will show how
to construct oscillating 
BPS solutions,
this section could be skiped in a quick reading.
Then we show how lower dimensional black holes 
are obtained from the ten dimensional  solutions. We discuss
the role of U-duality and  Dirac duality 
for  quantizing the charges. 
We finally consider extremal and non-extremal 
black hole solutions in five and four dimensions.
We will  define new variables identified with the number of   
some hypothetical   {\it non-interacting}
 ``constituents'' in terms of
 which the entropy takes a surprisingly simple form.

\subsec{Extended $p$-brane  solutions}

We will now consider solutions to type II  supergravity 
theories in ten dimensions. We will concentrate first with 
solutions that preserve some supersymmetries, 
the so called BPS solutions.
We start with one of the simplest, which  is the
solution corresponding to the fields outside 
a long fundamental string \dh . 
It only has fields in the first three terms in \actiontwoa\ excited
and it is  a solution in both type II theories  and also in the 
 heterotic 
 string theory. 
It carries charge under the NSNS $B_{\mu\nu}$ field,
this charge
appears as a central charge in the supersymmetry algebra.
The solution with the minimum mass for a given charge
will then be BPS.
The simplest way to find this BPS solution is the following.
Start with a SO(1,1)$\times$SO(8) symmetric ansatz for the
metric, in string frame,
\eqn\metrans{
ds^2 = h \left[ f_s^{-1} ( - dt^2 + dx_9^2 ) + dx_1^2 + 
\cdots + dx_8^2 \right]~.
}
We also allow the dilaton $\phi$ and the component $B_{09}$ of
the antisymmetric tensor to be nonzero and we set all 
other fields to zero. 
Now we try to find Killing spinors, 
which generate infinitesimal local  
supersymmetry
transformations that leave the solution invariant.
In order to be definite we consider the type IIA theory, 
a similar treatment goes through for the IIB and heterotic
theories.  The existence of unbroken supersymmetry 
implies that the gravitino and dilatino variations 
\eqn\susya{
\eqalign{
    \delta\lambda &= \left[ \p_\mu\phi\gamma^\mu\Gamma_{11}
        +{1\over 6}H_{\mn\rho}\gamma^{\mn\rho} \right] \eta ~,\cr
    \delta\psi_\mu &= \left[ \p_\mu + {1\over 4}\left(
        {\omega_\mu}^{ab}+{H_\mu}^{ab}\Gamma_{11}\right) \Gamma_{ab}
        \right] \eta , \cr}
}
should vanish for appropriate values of the spinor $\eta$, 
where $ \eta = \epsilon_R + \epsilon_L$ is the sum the 
a possitive and negative chirality spinor. 
Greek letters label coordinate indices, and latin letters
 label tangent space indices. Coordinate and tangent indices
are related by the zehnbeins $e^{a}_\mu$ and
$\omega_\mu^{ab} $ is the corresponding spin
connection.  $\Gamma_{a}$ are the flat space gamma matrices
satisfying $ \{ \Gamma_{a} , \Gamma_{ b} \} = 2
\eta_{ab}$, $ \gamma^{\mu } = e^\mu_{a}
\Gamma^{a} $ and $\gamma^{\mu_1\cdots \mu_n}$ is the
antisymmetrized product with unit weight (i.e. dividing by the number
of terms).
In order for the equations \susya\ to have solutions, the 
dilaton, the antisymmetric tensor field and the metric
have to be related to each other and take the form \dh
\eqn\fundstr{
\eqalign{
ds^2 =&  f_f^{-1}(- dt^2 
 + dx^2_9) + dx_1^2 + \cdots + dx_8^2~,
\cr
B_{09} =& {1 \over 2 }  (f_f^{-1} -1 )~,
\cr
e^{-2 (\phi -\phi_\infty ) } = & f_f ~,
}}
where $f_f$ is a function of the transverse coordinates
$x_1,...,x_8$ and  the rest of the fields are zero.
With this ansatz \fundstr\ the supersymmetry variations
\susya\ vanish if the spinors satisfy the conditions
\eqn\susyspinors{
\epsilon_{R,L} = f_f^{-1/4} \epsilon^0_{R,L} ~,~~~~~~~~~~~
\Gamma^0 \Gamma^9 \epsilon^0_{R} = \epsilon^0_{R}~,~~~~~~~~
\Gamma^0 \Gamma^9 \epsilon^0_{L} =- \epsilon^0_{L}~,
}
where the spinors $ \epsilon^0_{R,L}$ are independent of
position and are  the asymptotic values of the
Killing spinors. So we see that the solution 
preserves  half of the  supersymmetries for any function $f_f$. 
Actually, the  equations of motion of the theory (related to the
closure of the supersymmetry algebra) imply
that $f_f$ is a harmonic function $ \p^2 f_f =0$ where
$\p^2 $ is the flat Laplacian in the directions $1,\cdots, 8$. 
Taking 
\eqn\formofa{
f_f = 1 + { Q_f \over r^6} ~,
}
we get a solution that looks like a long string. It is
singular at $r=0$ but in fact one can see from the metric
that it is a so called null singularity, 
there is a horizon at the singularity and  we
do not have a naked singularity. 
In this classical solution the
constant $Q_f$ is arbitrary. However, this long string solution 
carries a charge under the $B$ field, this charge is carried
in string theory by the fundamental strings. The charge that
the fundamental strings carry is their winding number and it is
not continuous, it
is a multiple of  some minimum value. 
An easy way to see this is to consider this theory 
compactified on a circle by  periodically 
identifying  the direction $\hat 9 $ by $ x_9  \sim x_9 + 2 \pi R $. In
that case the $ B_{\mu 9} $ components of the antisymmetric tensor
field
 become a gauge field in the extended dimensions.
The ``electric'' charge associated with this gauge field
is the winding number along the direction $\hat 9$ which counts
how many strings are wound along this circle. In string theory 
this number is 
 an integer, there is a geometric quantization 
condition. This is why we say that 
 the fundamental strings can carry 
only integer multiples of this charge.
We conclude that $Q_f  = c_f^{10} m $, with $m$ an integer representing
the winding number. One can determine $c_f^{10}$ by comparing the 
charge of \fundstr\ with that of a fundamental string with winding
number $m$. This is equivalent, due to the fact that both are
BPS solutions, to comparing the masses. 
The ADM mass is determined from \fundstr\ from the $g_{00} $ 
component of the Einstein metric of the extended 
$1+8$ dimensional theory. 
This gives
\eqn\cf{
c_f^{10} = { 8 G_N^{10} \over \alpha' 6 \omega_7
} ~,}
where
$ \omega_d = { 2 \pi^{d/2} \over \Gamma(d/2) } $ is the
volume of the  sphere in $d$ dimensions $ S_{d-1} $.

Since this supergravity solution carries the same charge
and mass as the fundamental string and has the same
supersymmetry properties, it is natural to regard 
\fundstr \formofa \cf\ as describing the long range 
fields produced by a long fundamental string.
This is analogous to saying, in quantum electrodynamics,
 that the
electric field of a point charge describes the fields
far from an electron. 
Actually,
in \dh\ this coefficient \cf\ was determined by matching
the solution \fundstr\ to a fundamental string source
of the form \actionstr .

It is interesting that the equations of motion just demand
that $f_f$ in \fundstr\ is a harmonic function. Taking it
to be $ f_f = \sum_i c_f^{10}/(\vec r- \vec r_i)^6 $
 we describe a 
collection of strings sitting at positions $\vec r_i$ in
static equilibrium. The gravitational attraction and the 
dilation force cancel against the electric repulsion.
This ``superposition principle'' is a generic property 
of BPS solutions and will appear several times in
the construction of BPS black holes. 
We indeed expect to have no force since the energy of a 
BPS configuration with charge $m$, as given by the
BPS formula, does not depend on the position of the
charges.


Now we turn to other ten dimensional solutions that
preserve 1/2 of the supersymmetries. 
The fundamental string solutions carried 
``electric'' charge under the $B$ field.
 The corresponding field strength 
$ H = d B$ is dual to a seven index field strength
$ F_7 \sim   * H$ and can be written in terms of 
a six form $ F_7 = d \tilde{B}_6 $. This six form couples
naturally to a five-brane. 
The supergravity solution, called solitonic (symmetric) fivebrane, 
 is again determined in terms of a single 
harmonic function \callanfive .
 In string frame it reads 
\eqn\fivebrane{ \eqalign{
ds^2 = & -dt^2 + f_{s5} (dx_1^2 + \cdots + dx_4^2 ) + dx_5^2 + \cdots
+ dx_9^2 ~,\cr
e^{ - 2 (\phi-\phi_\infty )  } =& f_{s5}^{-1}~,
\cr
H_{ijk} =& (dB)_{ijk} = {1\over 2 } \epsilon_{ijkl} \p_l 
f_{s5} ~,~~~~~~~~~~~~~~~
i,j,k,l= 1,2,3,4~,
} }
and  all other fields are zero. $\epsilon_{ijkl}$  is just the 
flat space epsilon tensor. 
The  harmonic function $f_{s5}$  depends on the coordinates 
transverse to the fivebrane ($x_1 \cdots x_4$) and for
a single fivebrane takes the form
 $f_{s5} = 1 + { c_{s5} \over (x_1^2 +...+x_4^2)}
$.
The constant $c_{s5}$ is determined from the Dirac quantization 
condition. That is to say, the $B$ field that  results from
\fivebrane\ cannot be defined over all space and will have
have some discontinuities. These  discontinuities will be 
invisible to fundamental strings if the fivebrane
charge obeys the condition analogous to the Dirac 
quantization condition for electric and magnetic
charges. This  condition
implies that $c_5^s = \alpha' $ \callanfive , so
 that the mass of the fivebrane goes
as $ 1/g^2 $ showing a typical solitonic
behavior, what is more, the string  metric \fivebrane\ shows
a geometry with a long throat at $r=0$ so that it has
some ``size''. 
The Killing spinors that generate the unbroken supersymmetries
are determined, as in the case of the fundamental string 
\susyspinors ,
by some constant spinors at infinity which 
satisfy the conditions
\eqn\susyfive{
\epsilon^0_L =
 \Gamma^1\Gamma^2\Gamma^3\Gamma^4 \epsilon^0_L~,~~~~~~~~~~~~~
\epsilon^0_R = - \Gamma^1\Gamma^2\Gamma^3\Gamma^4 \epsilon^0_R~.
} 
Even though we have presented these solutions just
 as supergravity solutions it is possible to show that
they define conformal field theories, which implies that
they are solutions to the full string classical action, and
not just to the low energy supergravity. 


In type II theories it is natural to look for 
supergravity solutions describing the long range fields
away from  a D-brane. They will be extended branes of $p$ spatial 
dimensions, carrying ``electric'' charge under 
the $A_{p+1}$ forms, or ``magnetic'' under the $A_{7-p}$ 
forms.

These solutions  have the form,
in string frame \hs ,
\eqn\pbrane{
 \eqalign{
ds^2 = & f_p^{-1/2} ( - dt^2
+ dx_1^2 + \cdots + dx_p^2 ) + f_p^{1/2} (
dx_{p+1}^2 + \cdots + dx_9^2 )~,
\cr
e^{ - 2 \phi } =& f_p^{p-3 \over 2}~,
\cr
A_{0\cdots p} = & -{1 \over 2 }( f_p^{-1} -1 )~,
}}
where $f_p$ is again a harmonic function of the
transverse coordinates $x_{p+1},...,x_9 $. 
All these solutions are BPS and break half of the
supersymmetries through the conditions \susydbrane .
 They correspond to the 
extremal limit of charged black $p$-branes when
the harmonic function is $f_p = 1 + n c_p^{10}  /r^{7-p} $,
where $n$ is an integer and $c_p^{10}$ is related to the 
minimum charge of a D-brane and will be  calculated later using
U-duality.
In the type IIA we will have only solutions like
\pbrane\ for $p$ even and in the type IIB only for $p$ odd.
In  type IIB theory there are two kinds of strings:
the fundamental strings and the D-strings. Similarly 
there are two kinds of fivebranes, the solitonic
fivebrane and the D-fivebrane, the difference
between them is whether they carry charge
under the antisymmetric tensor field
$B_{\mu\nu}$ or $B'_{\mu\nu}$. The dilaton and the
string metric are also different in both solutions, but 
 they transform into each other under S duality.
The three brane is self dual under S-duality.

Note that all these extremal solutions are boost invariant
for boosts along the brane, in that sense they 
are relativistic branes like the fundamental string.
This property is related  to the
fact that they preserve some supersymmetries.
The extremal branes therefore cannot carry momentum in the
longitudinal directions by just moving in a rigid fashion
but, of course, they can carry transverse momentum. 
In order to carry longitudinal  momentum they  have to oscillate
in some way, that is the topic of the next section.
These oscillations propagate at the velocity of light since the
tension is equal to the mass per unit brane-volume.

\subsec{Oscillating strings and branes.}

This section is aimed at providing a more direct 
correspondence  between BPS oscillating strings 
and fundamental string states. It can be skipped 
in a first quick reading. 

As discussed in section 1.2  a  fundamental string 
containing only left moving oscillations is a BPS state
breaking 1/4  of the supersymmetries. It is natural
to look for supergravity solutions that describe the long distance
behaviour of these oscillating strings.  We
can take $R_9$ to be large and we  can make 
coherent states with the string oscillators, leading
to macroscopic classical oscillations. Therefore,  we expect 
the supergravity solutions to exhibit these oscillations which
describe traveling waves on a fundamental string.
The general method to construct these solutions was
developed by  \vespachati . In the case of fundamental
strings the oscillating solutions take the form \cmp\
\eqn\single{\eqalign{
ds^2 =& f_f^{-1}  du [ dv + \tilde k(r) du + 2 F'^i(u) dy^i ]
+ dy^i dy^i~,
\cr B_{uv} = & - {1 \over 4} ( f_f^{-1}-1) ~, \cr
B_{ui} = &   f_f^{-1}
F'^i(u) ~, \cr
e^{ -2 \phi } = & f_f~,
}
}
where $u=x_9-t,~~v=x_9 +t $ and 
 $F^i(u)$ are arbitrary functions describing a traveling wave on
the string. 
 $f_f$ and $\tilde k$ are harmonic functions. The solution
\single\ arises from the  chiral null models studied
in \hortsesing . 
  Since this metric is not manifestly asymptotically
flat, we prefer to make the simple change of coordinates
\eqn\change{
y^i= x^i - {F^i(u) } ~,~~~~~~~~~~v=\tilde v +   \int^u
{ \left[ F'^i(u_0) \right ]^2 du_0 ~. }
}
which puts the fields in the form
\eqn\singleasy{\eqalign{
ds^2 =& f_f^{-1}(\vec r , u) du \left[
d\tilde v -2 (f_f(\vec r,u) -1) F'^i(u) dx^i  + 
 k(\vec r,u) du \right] + dx^i dx^i  ~, \cr
k(\vec r, u) = & \tilde k(\vec r ,u) + ( f_f -1 ) \left( F'(u)
\right)^2 ~, \cr
B_{uv} = & - {1 \over 4} ( f_f^{-1}(\vec r,u) -1)~, \cr 
B_{ui} = &  \left( f_f^{-1}(\vec r,u) -1 \right) F'^i(u)~, 
}}
where $f_f(\vec r,u)=f_f(\vec r-\vec F(u))$ and $k(\vec r,u)= k(\vec
r-\vec F(u))$. Here  $f_f(r)$ is  as in
\formofa \cf\ with winding number $m$   and
$k(r)= P(u) {2 \pi \alpha' c_f^{10} /  r^6}
$, with $P(u)$ being the physical momentum per unit length 
 carried by
the string.
The metric is now manifestly asymptotically flat, and,
in the limit $F^i(u)\to 0$, it reduces to the static solution \fundstr .

This oscillating string solution \single\ preserves 1/4 of the 
supersymmetries. The spinors that generate the unbroken
supersymmetries satisfy
\eqn\susyosc{
\epsilon^0_R = \Gamma^0 \Gamma^9 \epsilon^0_R~,~~~~~~~~~~~~~~~~
\epsilon^0_L =0~.
}

As a  check on our understanding of the physics of these
solutions, we should verify that the excited strings do indeed
transport physical momentum and angular momentum. Since we have
written the metric in a gauge where it approaches the Minkowski metric
at spatial infinity, we can use standard ADM or Bondi mass techniques
to read off kinetic quantities from surface integrals over the
deviations of the metric from Minkowski form.  Following \dh \waldram
, we
pass to the physical (Einstein) metric $g_E = e^{-\phi/2} G_{string}$,
expand it at infinity as $g_{E \mu\nu} = \eta_{\mu\nu} + h_{\mu\nu}$
and use standard methods to construct conserved quantities from
surface integrals linear in $h_{\mu\nu}$. 
We find that the transverse momentum
per unit length on a slice of constant $u$ is
\eqn\mom{ P_i =  { m \over 2 \pi  \alpha'} F'^i(u) }
in precise accord with ``violin string'' intuition about the
kinematics of disturbances on strings. Similarly, the net
longitudinal/time energy-momentum per unit length 
$\Theta_{\alpha\beta}$, $\alpha, \beta = 0,9$,
 in a constant $u$
slice is
$$
(\Theta_{\alpha\beta}) =
\pmatrix{{ m\over 2 \pi \alpha'} 
+P(u) & -P(u) \cr -P(u) & - { m\over 2 \pi \alpha'} 
+P(u) \cr}~.
$$
 Finally, we
consider angular momenta. For the string in ten dimensions there are
four independent (spatial) planes and thus four independent angular
momenta $M^{ij}$ per unit length.
  Evaluating as an example $M^{12}$ we obtain \cmp\
\eqn\angularf{
M^{12} \sim ( f'^1 f^2 - f'^2 f^1)(u)~.
}
There are no surprises here, just a useful consistency check.

Note that a single fundamental string satisfies
the level matching condition \matching\ so we
might wonder if there is an analogous  condition
in the supergravity solution. One way to find this
condition is to demand that the solution matches to 
a fundamental string source \dabhargaunwald . 
Another way  is to demand that the
singularity, when we approach the string is not
naked but  null \cmp .
 This amounts to demanding that the 
function $\tilde k $ in \single\ vanishes,  
which leads
to 
\eqn\matching{
 { P(u) m \over 2 \pi \alpha }  = 
{  m^2 \over (2 \pi \alpha')^2 } F'^i(u)^2~.
}
There are also  BPS multiple string solutions where
the different strings are oscillating independently.
They are described in \cmp\  and
they involve new conformal field theories which 
are a generalization of the chiral null models
considered by \hortsesing . 
If we have such a  superposition 
the condition \matching\ need not be satisfied.
Actually one has  to effectively average over functions
$F^i(u)$ \cmp . For a general ensemble of functions 
$F'^i(u)$ will be uncorrelated with $F^j(u)$ and 
the $g_{\mu i},~B_{\mu i} $ components of the 
metric and antisymmetric tensor will vanish, leaving
just the function $k$ in \single . 
Note that this is not  the case if they are carrying
some net angular momentum \angularf .

\vskip 1cm
\vbox{
{\centerline{\epsfxsize=3in \epsfbox{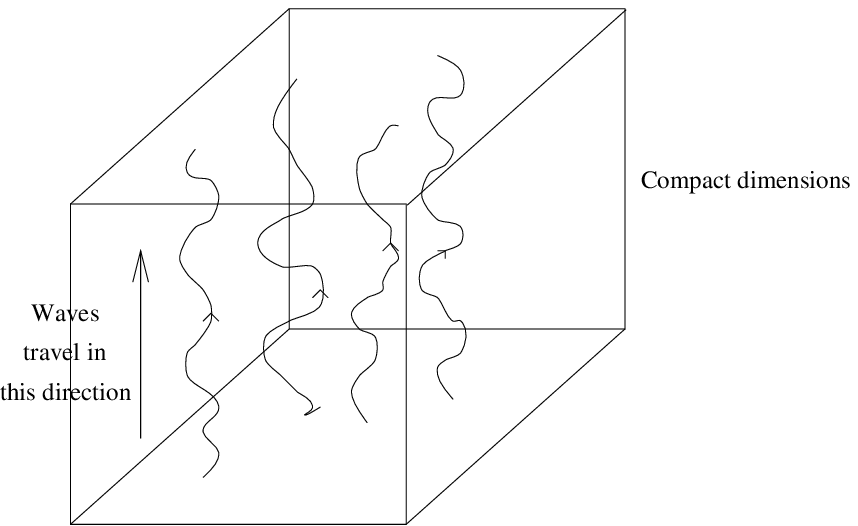}}}
{\centerline{\tenrm FIGURE 4:
Ensemble of many oscillating strings carrying 
traveling waves.
}}
}
\vskip .5cm

In a very similar fashion it is possible to construct 
oscillating $p$-branes.
In  fact, if we just average  over the oscillations
we simply get one more harmonic function 
$K= {c_P N \over r^{7-p}  } $ in the
solution. The coefficient $c_P = { \alpha' \over R^2_9 } c_f$
 is  calculated using U-duality (see section 2.4) and 
$N$ is the momentum, measured in units of the
minimum allowed. In conclusion, when 
 momentum is carried in a direction parallel to
the brane (call it $\hat 9$),
then the solution can be found
by replacing $-dt^2 + dx_9^2 \rightarrow -dt^2 + dx_9^2 +
k (dt - dx_9)^2 $ in the metric in \pbrane\ or 
\fivebrane\ . Adding momentum  leads to   BPS solutions 
preserving 1/4 of the supersymmetries by 
imposing the additional constraints  on the
spinors at infinity, due to the momentum, 
\eqn\susymom{
\epsilon_R^0 = \Gamma^0 \Gamma^9 \epsilon^0_R~,~~~~~~~~~  
\epsilon_L^0 = \Gamma^0 \Gamma^9 \epsilon^0_L~.
}


\subsec{$d\leq 9$ Black Holes From $d=10$ strings or branes.}

Since all the BPS solutions treated in the previous
section depend on some harmonic function $f$ one can 
make  multiple brane solutions by taking $f =
1 + \sum_i c_p/( \vec r- \vec r_i)^{7-p} $ which describes a
set of branes at positions $\vec r_i$ in static equilibrium. 
The gravitational attraction is balanced by the
repulsion due to their  charges. 

In this section the word ``brane'' will indicate 
any of the BPS solutions discussed above.
We will now consider the type IIB theory  compactified 
to $d$ dimensions on a torus $T^{10 -d}$, identifying
the coordinates by $ x_i \sim x_i + 2 \pi R_i $, choosing
periodic boundary conditions on this $ 10-d $ dimensional ``box''.
Fields that vary over the box will acquire masses of the 
order $m \sim 1/R $ where $R$ is the typical compactification
size. The easiest way
to see this is by expanding the fields in Fourier components
along the internal dimensions. So 
if we are interested in the low energy physics in 
$d$ extended dimensions  the fields will be independent
of the internal coordinates of the torus. 
If  we want to find solutions to this
$d$ dimensional supergravity theory, does it
help us to know the solutions in ten dimensions?
Yes, it does. The key point to observe is that if
we have any solution in ten dimensions which is 
periodic under $ x_i \rightarrow x_i + 2 \pi R_i $, then
it will also be  a solution of the compactified 
theory. For any $p$-brane, the solution is automatically 
translational invariant in the directions parallel to
the brane. In order to produce a periodic solution we 
superimpose BPS solutions forming a lattice 
in the transverse directions, producing  a harmonic function
\eqn\lattice{
 f = 1+ \sum_{\vec n \in Lattice} {  
c \over ( \vec r - 2 \pi R_i \vec n )^{7-p}  } ~.
}
We can view this as solving the Laplace equation with
the method of images in a periodic box. It is this nice
superposition principle for BPS solutions that enables
us to find a very direct correspondence between ten 
dimensional objects and the $d$ dimensional ones. 
We will be interested in solutions where  the 
brane is completely wrapped along the internal 
directions so that from the point of view of the
observer in  $d$ dimensions one has 
a localized, ``spherically'' symmetric solution.
These solutions will correspond to extremal limits
of charged  black hole solutions. 
The first point to notice that if the brane wraps
$p$ of the torus dimensions then the 
sum in \lattice\ runs over a $10-d-p$ dimensional lattice.
If we are looking at the solutions at distances much bigger than
the compactification scale, then we are allowed to replace
the sum in \lattice\ by an integral. This integral would naturally 
appear also if we average over the position of the
brane on the internal torus. The net effect of the 
integral will be to give the function $f = 1 + c_P^{(d)}/r^{d-3} $,
where $r$ is the distance in the extended $d$ dimensional
coordinates. Note that the power of $r$ is independent
of $p$ and is the appropriate one to be the
spherically symmetric solution to Laplace's equation
in $d-1$ spatial dimensions. 
So when we are in the $d$ dimensional theory, the only 
way we have to tell that the  black hole
contains a particular type of  $p$ brane is by looking at 
the gauge fields that it excites. The 
final result is that the $d$ dimensional solutions 
are given again 
by \fundstr , \fivebrane\ and \pbrane\ but  now in terms of 
$d$ dimensional  harmonic
functions. 

As a particular example we will consider the black holes
resulting from compactifying the oscillating strings
treated in the previous section. 
The oscillation will be along a compact direction, and
 we average over them. We could think
that we are looking at distances larger than
the compactification radius, or that we do an average over
the phase of the oscillation. It is important that this 
average is done at the level of the harmonic function
that specifies the solution, and not on the individual
components of the fields, which are non-linear in terms
of the harmonic functions. This procedure  
produces  a  solution of the $d$ dimensional supergravity
theory. 
To be more
precise, we build a periodic $(9-d)$ dimensional array of strings by
taking the harmonic function as in \lattice .
For large distances $\rho$ in the extended dimensions 
we can ignore the dependence on the internal
dimensions and find, 
\eqn\lambdaddim{
f_f^{(d)} = 1+ {c^{(d)}_f  m \over \rho^{d-3}}~,~~~~~ 
\qquad {\rm where}\quad
  c^{(d)}_f  = {{16\pi G_N^d R_9 }\over{\alpha' (d-3)\omega_{d-2} }}~,
}
and $\omega_d$ is the area of the $d$ dimensional unit sphere and
$m$ is the total winding number. We
could have taken directly $f_f^{(d)}$ as a solution of
the Laplace equation in the uncompactified dimensions, but we 
obtained it from superimposing solutions to clarify the
 connection to underlying string states. As we
will now show, the result of this procedure can be interpreted as a
lower-dimensional extremal black hole. The general idea that
ten-dimensional string solutions can be used to generate
four-dimensional black holes is not new and has been explored in
\hortseprl ,\ber ,\kallosh ,\mirjam ,\cmp .

We now look in more detail at the $d$-dimensional fields generated by
this compactification. Using the dimensional reduction procedure of
\sm\ , we find that the
$d$-dimensional fields obtained from wrapping a string with oscillations
are, in $d$-dimensional Einstein metric,  
\eqn\ddimdil{\eqalign{
e^{-2 \phi_d } = & e^{-2\phi_{10}} \sqrt{ G_{99}}
= \sqrt{  f_f^{(d)} ( 1 + k^{{d}} ) }
\cr
ds^2_E =& - { 1 \over \left[  f_f^{(d)} ( 1 + k^{{d}} ) 
\right]^{ d-3 \over d-2 } } dt^2 + 
\left[ f_f^{(d)} ( 1 + k^{{d}} ) \right]^{ 1 \over d-2 } d\vec x ^2~.
}}
This and all the other fields obtained by dimensional reduction turn
out to be the type II analogs   of Sen's four-dimensional black holes and
their higher-dimensional generalizations 
\senbh , \peet .
The Einstein metric of the $d$ dimensional solution has the same
form if we consider  any  other oscillating brane completely 
wrapped around  the internal torus since the Einstein metric  is
 invariant under U-duality. 
We can check that for these black  holes \ddimdil\ the 
area of the horizon, which is at $\rho =0 $,  is zero, so that
the classical entropy  is zero. It is possible to define
a nonzero ``classical'' entropy at the ``stretched'' horizon
which agrees up to a numerical constant with the counting
of states \senbhstr .


\subsec{U-duality and quantization of the charges}

We will show in this section how to quantize the 
charges using U-duality \hull . There has been some disagreement
in the literature concerning the precise quantization condition 
so  we have decided, for completeness, to explain it in detail.
Since the quantum of charge
will depend on the normalization chosen for the gauge 
field we find it more convenient to find the 
``quantum of mass''. This quantity has a well defined
meaning  since the solution is BPS and the mass is proportional
to the charge and 
protected from quantum corrections so that it can
be calculated using the weakly coupled theory. 
When we perform S-duality transformations
we should remember that the mass measured in 
the Einstein metric $ g_E = e^{ - { \phi /2 } } G $
(which includes a power of $g$) stays invariant. This is not 
how  we normally measure masses,
 we normally leave a power of $g$
in the Newton constant. 
 The masses we are going to calculate
are defined in terms of a modified Einstein metric which
is $ \tilde g_E = e^{ - { ( \phi -\phi_\infty )\over 2   } }
G = g^{1/2} g_E $ which agrees with the string metric at 
infinity. All we are saying is 
 that we keep the factor of $g^2$ in the Newton
constant.  Masses measured in the two metrics differ by
$M_E = g^{1/4} M $, where $M$ is the mass measured 
in the metric $\tilde g_E $ which is the one we are
going to use here. 
 The $d$ dimensional Newton
constant is $G_N^d = G_N^{10}/V_{10-d}$ where
$V_{10-d}$ is the volume of the internal torus.
We start with the minimum mass of a winding string which is
\virasoro\ 
\eqn\massfund{ M_{f} = { R_9 \over \alpha'} ~. }
Similarly the minimum mass for momentum states is $M = 1/R $.
Now we want to calculate the mass of a D-string with unit
winding using ten dimensional S-duality.
 We know that the Einstein metric is
invariant under S-duality so that $M_E$ is invariant,
this implies 
\eqn\massdf{g'^{1/4} M'=
M_E' = M_E = g^{1/4} {R_9\over \alpha' } 
}so that
the mass  of the D-string is
\eqn\massdstr{
M^{1D} = { R_9 \over g  \alpha' }~,
}
where we took into account the change in $R_9$ as in \sduality .
Applying T-duality transformations \tduality\ along a direction
perpendicular to the D-string we turn it into a D-twobrane
with mass
\eqn\massdtwo{ M = {R_9 \over g \alpha' } = { R_9 R'_8
\over g' \alpha'^{3/2}  }~.
}
Proceeding in this fashion we 
find the minimum mass for any D-brane
\eqn\massdbrane{
M^{pD} = { R_{10-p} \cdots R_9 \over g \alpha'^{(p+1)/ 2 } }~.
}
Doing now an S-duality transformation 
on the D-fivebrane, as in \massdf\  we get the mass
of the solitonic fivebrane
\eqn\massfive{
M^{s5} = { R_5 \cdots R_9 \over g^2 \alpha'^{ 3 }
}~.}
Our objective is to determine the coefficients that
appear in the harmonic functions specifying the solutions
\fundstr \fivebrane \pbrane . 
Since we will be mainly interested in four and five
dimensional black holes we are interested in the
coefficient that appears in the $d$ dimensional 
harmonic functions as in \lambdaddim \ddimdil . Actually
from \ddimdil\ by  setting  $k^{(d)} = 0$ we can read 
off the mass of these objects in terms of the coefficients
in the harmonic function $f^{(d)}$. The mass is 
calculated  from the behaviour of $\tilde g_{E \, 00}$ of the
metric at infinity \myresperry\
\eqn\adm{
\tilde g_{E \,00} \sim {
16 \pi G_N^d  M  \over (d-2)\omega_{d-2} }{1 \over r^{d-3} } =
 { d-3 \over d-2 } 
  {  c^{(d)} \over r^{d-3}  }
}
where $\omega_n$ is the volume of the unit sphere $S_n$, $
\omega_n = { 2 \pi^{n/2} \over \Gamma(n/2) } $.
This determines the coefficients for all excitations.
We still have to express $G_N $ in terms of $g$, remember that
we defined $g$ to be such that it goes to $1/g$ under S-duality
 \sduality .  
In order to do that, we use Dirac duality of the 
fundamental string and the solitonic fivebrane. 
The fundamental string carries electric charge under the NSNS
$B_{\mu\nu}$ field while the solitonic fivebrane carries magnetic
charge. It is not possible to define globally the $B_{\mu \nu}$
field of the fivebrane \fivebrane . This field will contain 
a singularity, analogous to the Dirac string for a monopole
in electrodynamics. The condition that this 
 singularity is  invisible for the
fundamental strings
fixes the coefficient of the fivebrane
harmonic function as $ c_{s5}^{(5)} = \alpha' $ \callanfive \foot{
Note that in comparing with \callanfive\ we only have to 
check that they used the same definition of the string
tension as in \actionstr .}.
Comparing this value with the one resulting from \adm\ and
\massfive\ we find the ten dimensional Newton constant
\eqn\gten{
G_N^{10} = 8 \pi^6 g^2 \alpha'^4~.
}

In string theory one can independently calculate the 
mass of D-branes from virtual closed string exchange diagrams,
in a similar fashion as one calculates the force between 
two charges in quantum electrodynamics.
The string ``miracle''  \polchinski\ 
is that this  string theory calculation  of
 masses of D-branes
agrees  with the masses predicted  by 
 U-duality as above.

Now for later convenience let us quote the results, which 
are obtained from \massdbrane \adm \gten\
 for
the D-onebrane, D-fivebranes and momentum in five extended
dimensions, which we will need for the five dimensional
black holes,
\eqn\constants{
c^{(5)}_1 = { 4 G_N^5  R_9 \over \pi \alpha' g }~,~~~~~~~~~~
c^{(5)}_5 = { g \alpha'  }~,~~~~~~~~
c^{(5)}_P = { 4 G_N^5  \over \pi  R_9  }~.
}

We will also need the corresponding coefficients
for D-twobranes, D-sixbranes, solitonic fivebranes
and momentum in four extended dimensions
\eqn\constfour{\eqalign{
c^{(4)}_2 = &{ 4 G_N^4 R_4 R_9 \over g \alpha'^{3/2} }~,~~~~~~~~
c^{(4)}_5 ={  \alpha'  \over 2   R_4  }~, \cr
c^{(4)}_6 = &{ g \alpha'^{1/2}  \over 2  }~,~~~~~~~~~~~~~
c^{(4)}_P = { 4 G_N^4 \over  R_9 }~,
}}
where we have used the value of the Newton constant \gten . 

\subsec{Black hole solutions in five dimensions.}

In section 2.3  we considered black holes 
coming from wrapping just one type of branes on the
torus, or at most one type of branes with oscillations. 
All  those black holes have zero horizon area 
and are singular at the horizon, since there are
scalar fields diverging at the horizon.
By looking at the classical solutions we see that
in almost all of them the dilaton is going to plus or
minus infinity.
 Also the physical longitudinal size goes to zero, 
measured in Einstein metric, for  all the branes with no oscillations.
Adding momentum in the internal directions does not 
help, we still have some diverging scalar. 
The three brane \pbrane\ has constant dilaton but suffers
of this problem about the physical size.

Our goal is to construct solutions with well defined 
geometries at the horizon, like the ones appearing in 
General Relativity. 
The key principle is that we need to balance the scalars
at the horizon. Different branes have different scalar charges,
which can be interpreted as pressures or tensions in the
compact direction. Note that even the dilaton falls in this
category when we think of it as the size of the 11$^{th}$ 
dimension in M-theory. 
If a scalar diverges when we approach the horizon the
$d$ dimensional character of the solution is  lost. 
This forces us to consider more than one type of 
branes. We need three different types for black holes
in five dimensions and four different types for
black holes in $d=4$, here two non-parallel $p$-branes
count as being of different type.

\vskip .1in
\noindent
{\it 2.5.1. Extremal black holes in five dimensions.}
\vskip .1in

We construct the five dimensional black hole with
nonzero area by superposing a number $Q_5$ of D-fivebranes,
$Q_1$ D-onebranes and Kaluza-Klein momentum.  
We consider type IIB  compactified 
 on $T^5$. We wrap a number $Q_5$ of  
D-fivebranes  on $T^5$. Then we wrap $Q_1$ D-strings
along one of the directions of the torus, let us pick the
9$^{th}$ direction.
In addition we put some momentum $P_9 = N/R_9$  along the 
string, i.e.  in the direction $\hat 9$.
The solution is given by three harmonic functions
$f_5, ~f_1$ and $k$. 
We start  writing  the solution in terms of the ten 
dimensional string metric, so that the relation
to \pbrane\ becomes more apparent \tseytlin \cama\
\eqn\tenfived{
\eqalign{ d s^2_{str} = &  f^{-{1\over 2}}_1
f^{-{1\over 2}}_5
 \left( -dt^2 + dx_9^2 +
k (dt-dx_9)^2 \right) + \cr
 +&f^{{1\over 2}}_1 f^{{1\over 2}}_5 ( dx_1^2 +  \cdots + dx^2_4) +
  f^{{1\over 2}}_1 f^{-{1\over 2}}_5 ( dx_5^2  + \cdots + dx^2_8 )~,
\cr 
e^{- 2 (\phi_{10}-\phi_\infty) } =  & { f_5 ~ f_1^{-1} }~,
\cr 
B'_{09} = & { 1\over 2} ( f_1^{-1} -1 )~, \cr
H'_{ijk} = & ( d B')_{ijk} = {1\over 2 }
\epsilon_{ijkl} \p_l f_5 ~,~~~~~~~~i,j,k,l = 1,2,3,4
}}
where $\epsilon_{ijkl}$ is again the flat space epsilon tensor. 
The three harmonic functions are
\eqn\harmonic{ f_1 = 1 + { c_1^{(5)} Q_1 \over x^2 }~,~~~~~~~~~~~~~~~
f_5 = 1 + { c_5^{(5)} Q_5 \over x^2 } ~,~~~~~~~~~~~~~~
k =  { c_P^{(5)} N  \over x^2 }
}
with $x^2 = x_1^2 +  \cdots + x^2_4$ and the coefficients
are given in \constants . 
The components of the Ramond-Ramond antisymmetric
tensor field, $B'^{RR}_{\mu\nu}$,
that are  excited    behave as gauge fields when
we dimensionally reduce to five dimensions.
The three independent charges arise as follows:
$Q_1$ is a RR electric charge, coming from $B'^{RR}_{09}$ and
counts the 1D-branes.
 $Q_5$ is a magnetic charge for the
three form field strength $H'^{RR}_3 = d B'^{RR}_2 $, which is
dual in five dimensions to a gauge field, $ F_2 = *_5 H'^{RR}_3$.
 $Q_5$ is thus an
electric charge for the gauge field $F_2$ and it counts the
number of 5D-branes.
The third charge, $N$, corresponds to
 the total momentum along the branes
in the direction $\hat 9$,
and  it is associated
to the five dimensional
Kaluza-Klein
gauge field coming from the $G_{09}$ component
of the metric.

Let us understand what happens to the
supersymmetries.
In  the ten dimensional type IIB theory the supersymmetries
are generated by  two independent chiral spinors $\epsilon_R$
and $\epsilon_L$ ( $\Gamma^{11} \epsilon_{R,L} =\epsilon_{R,L}$).
The presence  
of the D-strings and the D-fivebranes imposes  additional conditions
on the surviving supersymmetries
\eqn\susyonefive{
 \epsilon_{R} = \Gamma^0 \Gamma^9 \epsilon_L ~,~~~~~~~~
 \epsilon_{R} = \Gamma^0 \Gamma^5 \Gamma^6\Gamma^7\Gamma^8\Gamma^9
 \epsilon_L~,
}
where the first condition is due to  the presence of the
string and the second to the presence of the fivebrane \susydbrane .
When we put momentum we break additional supersymmetries through
the conditions 
\eqn\extra{
\Gamma^0 \Gamma^9 \epsilon_R =  \epsilon_R~,~~~~~~~~~~
\Gamma^0 \Gamma^9 \epsilon_L =  \epsilon_L .
}
Taken together with \susyonefive\ we get the following 
decomposition of the spinor under the group
 S0(1,1)$\times$SO(4)$_E \times$SO(4)$_I$, which is the subgroup 
 of Lorentz transformations that leaves the
ten dimensional 
solution \tenfived\ invariant, 
\eqn\spinor{
\epsilon_L = \epsilon_R=
\epsilon_{SO(1,1)}^+ \epsilon_{SO(4)}^+ \epsilon_{SO(4)}^+~.
}
The positive chirality SO(4) spinor is pseudoreal and has two independent
components so that 1/8 (4 out of the original 32) 
 supersymmetries are preserved
by this configuration.
The first  SO(4)$_E$
corresponds
to spatial rotations in 4+1 dimensions.  SO(4)$_I$
 corresponds
to rotations in the internal directions $5,6,7,8$ and is
broken by the compactification. 
The solution is supersymmetric, and has the same energy, 
independent on whether all the branes are sitting at the
same point or not, so in principle we can separate the
different constituents of the black hole. The resulting
black hole will have lower entropy so this process 
violates the second law of thermodynamics.

Now we dimensionally reduce \tenfived\ to five dimensions in order to
read off black hole properties.
The standard method of  \dimensionalreduction ~yields
a five-dimensional Einstein metric,
$ g^5_E = e^{ - 4 \phi_5/3 } G^5_{string} $,
\eqn\einstein{
ds^2_E = - { 1 \over
\left( f_1 f_5 (1+k) \right)^{2 \over 3} } dt^2 +
\left( f_1 f_5 (1+k) \right)^{1 \over 3}
( dx_1^2 + \cdots + dx_4^2 )~,
}
which describes a five dimensional extremal, charged,
supersymmetric black hole with nonzero horizon area. Calculating the
horizon area in this metric \einstein ~we get the entropy
\eqn\entropyext{
S_e  =
 {A_H \over 4 G^5_N } =  2 \pi \sqrt{N Q_1 Q_5 }~.
}
In this form the entropy does not depend on any of the continuous
parameters like the coupling constant or the sizes of the internal
circles, etc. This ``topological'' character of
the entropy was emphasized in  
\topological , \larsen  , \tseytlin .
 It is also symmetric under interchange of
$N, Q_1, Q_5$. In fact, U duality  
\hull , \sentduality ,
\vafaintersecting\ 
interchanges the three  charges.
 To show it in a more specific fashion,
 let us define T$_i$ to be the usual T-duality that inverts the
 compactification radius in the direction $i$
 and S the ten dimensional S duality of
 type IIB theory. Then a transformation that
sends ($N,Q_1, Q_5$) to ($Q_1, Q_5, N$)
 is U= T$_8$T$_7$T$_6$T$_5$ST$_6$T$_9$. 
Note however that this transformation changes the 
coupling constant and the sizes of the $T^5$.

The standard five-dimensional extremal \RN\ solution \myresperry\  
 is recovered
when the charges are chosen such that
\eqn\balance{
c_P N = c_1 Q_1=c_5 Q_5 =r_e^2~.
}
The crucial point is that, for this ratio of charges, the dilaton field
and the internal compactification geometry are independent of position
and the distinction between the ten-dimensional and five-dimensional
geometries evaporates. What is at issue is not so much the charges
as the different types of energy-momentum densities with which they are
associated. 
An intuitive picture of what goes on is this \cama : a p-brane produces
a dilaton field of the form $ e^{- 2 \phi_{10} } = f_p^{p-3 \over 2}$,
with $f_p$ a harmonic function \hs. A superposition of branes produces
a product of such functions and one sees how 1-branes can cancel
5-branes in their effect on the dilaton. A similar thing is true for
the compactification volume: For any p-brane, the string metric is such
that as we get closer to the brane the volume parallel to the
 brane shrinks, due to the brane tension,  and the volume perpendicular
to it expands, due to the pressure of the electric field lines.
 It is easy to see how
superposing 1-branes and 5-branes can stabilize the volume in the directions
$ \hat 6, \hat 7, \hat 8, \hat 9$, since they are 
 perpendicular to the 1-brane 
and
parallel to the 5-brane. The volume in the direction $\hat 5$ would still
seem to shrink, due to the tension of the branes. This is indeed why
we put momentum along the 1-branes, to balance the tension and produce a 
stable
radius in the $\hat 5$ direction. If we balance the charges precisely
\balance ~(we can always do
this for large charges) the moduli scalar fields associated
with the compactified dimensions are not excited at all, which is what 
we need
to get the \RN\ black hole.
Of course, if we do not balance them precisely we still have
a black hole with nonzero area, as long as the three charges 
are nonzero.


\vskip .1in 
\noindent
{\it 2.5.2 Non-extremal black holes in  five dimensions.}
\vskip .1in

The five dimensional \RN\ black hole is a solution of
the five dimensional Einstein plus Maxwell action.
The metric reads \myresperry
\eqn\reno{
ds^2 = - \lambda dt^2 + \lambda^{-1} dr^2 + r^2 d\Omega_3^2 ~,}
$$ \lambda = \left(  1-  {r_+^2 \over r^2}  \right)
\left(  1-  {r_-^2 \over r^2}  \right)~.
$$
There is a horizon at
 $ r = r_+ $,
mass and charge are given by
\eqn\mass{
 M = { 3 \pi \over 8 G^5_N } ( r_+^2 + r_-^2) ~,~~~~~~~~~
 Q =   { 3 \pi \over 4  G^5_N }  r_+  r_- ~.
}
The extremal solution is obtained by taking $r_+ = r_- \equiv r_e $
and reduces to \einstein , with the charges related by
\balance , after doing the 
coordinate transformation $r^2 = x^2 + r_e^2$.

Now we would like to construct the non-extremal five dimensional
black holes with arbitrary values of the charges.
The method is very simple \hstro \hms . First we start with the 
non-extremal \RN\ \reno\ which has some constraints on the 
charges \balance , then we lift up this configuration to ten dimensions.
That is done by inverting the standard dimensional
reduction procedure \dimensionalreduction ,
and we  find the ten dimensional form of the various 
fields. This gives a non-extremal configuration where
the charges are related by  \balance . We will apply some
transformations which remove the constraints of \balance .
We start by  boosting  the solution along the direction of 
the onebranes  (we  called it $\hat 9$).
 This introduces some
extra momentum, so that now  the RR charges are constrained but 
the momentum is arbitrary. The   result is a solution 
which can be viewed  as a   black string in six dimensions   \hstro  .
Now we need to remove the constraint on the RR charges.
To that effect   we do 
a U  duality transformation  that interchanges the 
three different charges. More precisely we perform 
the transformation 
 U=T$_8$T$_7$T$_6$T$_5$ST$_6$T$_9$
that  sends ($N,Q_1, Q_5$) to ($Q_1, Q_5, N$).
This transformed one RR charge into momentum, so that
we can boost the solution to  produce a solution
with arbitrary  value of this RR charge.
After doing all these transformations, and choosing 
some appropriate coordinates,
the resulting  ten dimensional solution is, in string
metric,
\eqn\dil{ e^{-2( \phi -\phi_\infty) } = 
 \(1+   { \sg \over r^2 }\)\(1 + {\sa\over r^2 }\)^{-1}~,
}
\eqn\metric{\eqalign{
ds^2_{str} = &
 \( 1 + { \sa \over r^2}\)^{-1/2} \( 1 + { \sg \over r^2}\)^{-1/2}
\left[ - dt^2 +dx_9^2
\right. \cr
+& \left. {
r^2_0  \over r^2} (\cosh \sigma dt + \sinh\sigma dx_9)^2
 +\( 1 + {\sa \over r^2}\) (dx_5^2 + \dots + dx^2_8) \] \cr
 +& \( 1 + { \sa \over r^2}\)^{1/2}\( 1 + { \sg \over r^2}\)^{1/2} 
\left[
\(1-{r_0^2 \over r^2}\)^{-1} dr^2 + r^2 d \Omega_3^2 \right]~.
}}
This solution is parameterized by the six independent quantities
$\alpha,\g,\sigma,r_0,R_9\equiv R $ and $V$. The last two
parameters are the radius of the 9$^{th}$ 
 dimension and the product
of the radii in the other four compact directions $V = R_5 R_6 R_7
R_8 $. They appear in the charge quantization 
conditions, indeed the three charges are
\eqn\charges{
\eqalign{
   Q_1 &= {V\over 4\pi^2 g}\int e^{\phi_6} *H'
   = { V r_0^2  \over 2 g } \sinh 2 \a , \cr
   Q_5 &= {1\over 4\pi^2 g} \int H'  =  { r_0^2\over 2g} \sinh 2 \g ,
\cr
  N &= {  R^2V r_0^2 \over 2  g^2} \sinh 2 \sigma ,
}}
where $*$ is the Hodge dual in the six dimensions $x^0,..,x^5$.
For simplicity we  set from now on $\alpha'=1$. 
The last charge $N$ is related to the  momentum around the $S^1$
by $P_9=  N/R_9$. All charges are normalized to be integers.

Reducing \metric\  to five dimensions using \dimensionalreduction ,
 the solution takes
the remarkably simple and symmetric form:
\eqn\solnfd{ds_5^2 =  - \lambda^{-2/3} \(1-{r_0^2 \over r^2}\) dt^2 + 
\lambda^{1/3}
\[\(1-{r_0^2 \over r^2}\)^{-1} dr^2 + r^2 d \Omega_3^2 \right]~,}
where
\eqn\deff{ \lambda 
= \(1+{\sa\over r^2} \)\(1+{\sg\over r^2} \)\(1+{\ss\over r^2} \)~.}
This is just the five-dimensional Schwarzschild metric with the time
and space components rescaled by different powers of $\lambda$. 
The factored
form of $\lambda$
 was known  to hold for extremal solutions \einstein\  \mirjam . 
It is surprising that
it continues to hold even in the non-extremal case. The solution is
manifestly invariant under permutations of the three boost parameters as
required by U-duality.
The event horizon is clearly at $r=r_0$.
The coordinates we have used present the solution
in a simple and symmetric form, but they do not always cover the entire
spacetime. When all three charges are nonzero, the surface $r=0$ is
a smooth inner horizon. This is analogous to the situation in four
dimensions with four charges \cvetd  .
 When at least one of the charges
is zero, the surface $r=0$ becomes singular.
Several thermodynamic quantities can be associated to
this solution. They can be computed in either the ten dimensional or
five dimensional metrics and yield the same answer. For example, the
ADM energy is 
\eqn\mss{E=  {   R V r_0^2  \over 2  g^2}
(\cosh 2 \alpha + \cosh 2 \gamma + \cosh 2 \sigma  )~.
}
The Bekenstein-Hawking entropy is
\eqn\entropy{
S = {A_{10}\over 4 G^{10}_N} = {A_5\over 4 G^5_N} =
{ 2 \pi  R V  r_0^3 \over  g^2 } 
 \cosh \alpha \cosh \gamma \cosh \sigma.
}
where $A$ is the area of the horizon and
we have used the value \gten\ for the Newton constant.
The Hawking temperature is
\eqn\thwk{T= {1 \over 2 \pi  r_0 \cosh\alpha \cosh \gamma \cosh \sigma } .}
In ten dimensions, the black hole is characterized by
pressures which describe how the energy changes for isentropic
variations in $R$ and $V$. In five dimensions, these are `charges' 
associated
with the two scalar fields coming from the components
$G_{99}$ and $G_{55}$ in \metric , which 
can be interpreted as the pressures in the directions 
9 and 5 respectively, and they read
\eqn\psses{
\eqalign{
P_1=&
{ R V r_0^2 \over 2  g^2} \left[ \cosh 2 \sigma -  {1\over 2} (
\cosh 2 \a + \cosh 2 \g ) \right]~,
\cr
P_2 =&
{ R V r_0^2\over 2  g^2} ( \cosh  2 \a -\cosh 2\g )~.
}}

The extremal limit corresponds to the limit $ r_0 \rightarrow 0$
with at least one of the boost parameters
$\alpha , \gamma , \sigma \rightarrow \pm \infty $ keeping
$R, \ V$ and the associated charges \charges\ fixed.
If we keep all three charges nonzero in this limit, one obtains
\eqn\extl{\eqalign{E_{ext} &= {R |Q_1|\over  g}  +
{  R V |Q_5|\over  g}  +
 { |N|  \over R}  ~,\cr
S_{ext}&=2\pi\sqrt{|Q_1Q_5 N|}~,\cr
            T_{ext}&=0~,\cr
P_{1ext}&= { |N| \over R }  - { R |Q_1|\over 2  g } -
 {R V |Q_5| \over 2 g }~, \cr
P_{2ext}& = {R |Q_1 | \over  g}  - { R V |Q_5|\over  g }  ~.}}
The first equation is the saturated Bogomolnyi bound for this theory.

We now show that there is a formal sense in
which the entire family of solutions
discussed above  can
be viewed as  ``built up" of branes, anti-branes, and momentum.
The extremal limits with only one type of excitation are
 obtained
by letting $r_0$ go to zero and  taking a   
boost parameter go to infinity keeping only one
charge  fixed. These extremal metrics represent a D-onebrane wrapping
the $S^1$, or  a D-fivebrane wrapping the $T^5$,
or the momentum modes
around the $S^1$.
 From \mss\ and \psses\ we see that a single
 onebrane {\it or} anti-onebrane has
mass and pressures
\eqn\obc{M={R\over  g}, \qquad P_1 = -{R\over 2g}, \qquad P_2 = {R\over g}.}
Of course a onebrane has $Q_1 =1$, while an anti-onebrane has $Q_1 = -1$.
A single fivebrane  or
anti-fivebrane has
\eqn\fbc{M={RV\over  g}, 
\qquad P_1 = -{RV\over 2g} ,\qquad P_2 = - {RV\over g }.}
For left- or right-moving momentum
\eqn\sbc{ M={1\over  R}, \qquad P_1 = {1\over R} ,\qquad P_2 = 0}

Given \obc\ - \sbc,  and the relations \charges, \mss, and \psses,
it is possible to trade the six parameters of the general solution
for the six quantities ($N_1,~N_{\bar 1},~N_5,~N_{\bar 5},~N_R,~N_L$)
which are the ``numbers'' of onebranes, anti-onebranes, fivebranes,
anti-fivebranes, right-moving momentum and left-moving momentum
respectively. This is accomplished by equating the total
mass, pressures and charges of the black hole with
those of a collection of  ($N_1,~N_{\bar 1},~N_5,~N_{\bar 5},~N_R,~N_L$)
{\it non-interacting} ``constituent'' branes, antibranes and momentum.
By non-interacting we mean that the masses and pressures are simply
 the sums
of the masses and pressures of the constituents.
The resulting expression for the $N$'s are
\eqn\dbranes{
\eqalign{
   N_1=& { V  r_0^2  \over 4g} e^{ 2 \a},
\cr
    N_{\bar 1}= &
{ V  r_0^2  \over 4 g} e^{- 2 \a}~,
\cr
 N_5 =& { r_0^2 \over 4g} e^{ 2 \g },
\cr
 N_{\bar 5} =&
{ r_0^2 \over 4g} e^{- 2 \g },
\cr
 N_R = &
{   r_0^2 R^2 V  \over 4 g^2} e^{ 2 \sigma },
\cr
 N_L = & { r_0^2 R^2 V  \over 4 g^2} e^{ -2 \sigma }.
}
}
\dbranes\ is the {\it definition} of the $N$'s, but we will refer to them
as the numbers of branes, antibranes and momentum  because
(as will be seen) they reduce
to those numbers in certain limits where these concepts are well defined.

In terms of the numbers \dbranes, the charges are simply
$Q_1 = N_1 - N_{\bar 1}, \ Q_5 = N_5 - N_{\bar 5}, \ N = N_R - N_L$,
the total energy is
\eqn\energybr{ E = {R\over  g} (N_1 + N_{\bar 1}) + { RV \over  g}
     (N_5 + N_{\bar 5}) + {1 \over R} (N_R + N_L)~,   }
and the volume  and radius are
\eqn\forvol{ V =  \( {N_1 N_{\bar 1} \over
N_5 N_{\bar 5}} \)^{1/2}~,}
\eqn\forlen{ R =  \( { g^2 N_R N_L \over N_1 N_{\bar 1} } \)^{1/4}~.
}
 From \extl\ we see that the extremal solutions correspond to including
 either branes or anti-branes, but not both. Notice that for the general
 Reissner-Nordstrom solutions ($\a = \g = \sigma$) the contribution to
 the total energy from onebranes, fivebranes, and momentum  are all equal:
\eqn\rnequal{ {R\over  g} (N_1 + N_{\bar 1})={ RV \over  g}
(N_5 + N_{\bar 5})={1 \over R} (N_R + N_L)   .}
The actual number of branes of each type  depends on $R$ and $V$ and
can be very different.

Of course there seems to be no reason for neglecting interactions
between collections of branes and momentum modes
composing a highly non-extremal black hole at strong or intermediate
coupling. Hence
the definitions \dbranes\ would seem to be inappropriate for
describing a generic black hole.
However, the utility of these definitions can be seen when we
reexpress the black hole entropy \entropy\ in terms of the $N$'s.
It takes the remarkably simple form
 \eqn\smira{S= 2 \pi( \sqrt{ N_1} + \sqrt{  N_{ \bar 1}})
( \sqrt{ N_5} + \sqrt{  N_{ \bar 5}})( \sqrt{ N_L} + \sqrt{ N_R})~.}
In the next chapter we will compute this formula in string
theory in  some special limits. An interesting property of this 
entropy formula \smira\ ~is that if one takes the brane-antibrane
numbers to be free variables and then one maximases the entropy
\smira\ ~subject to the constraints that the charges and the 
 total energy
 \energybr\ ~are fixed, then one gets the relations 
\forlen \forvol\ ~and hence \dbranes\ ~for the brane-antibrane
numbers. So, in this very specific  sense, 
the black hole solution represents
a system of branes and antibranes in thermodynamic equilibrium.

A puzzling feature of \smira\ is that it only involves onebranes,
fivebranes, and momentum. This is understandable for extremal solutions
with these charges, but when one moves away from extremality, one might
expect pairs of threebranes and anti-threebranes or fundamental string
winding modes to contribute to the entropy. To 
understand the roles
of these other objects, one should start with the full Type II 
string theory
compactified on $T^5$. The  low energy limit of this theory
is $N=8$ supergravity
in five dimensions (we measure $N$ in four $d$ terms,
i.e. by the amount of supersymmetry that it has reduced to $d=4$).
 This theory has 27 gauge fields, 42 scalars and
a global $E_6$ symmetry. Since only the scalar fields which couple to the
gauge fields
 are nontrivial in a black hole background, we expect the general
solution to be characterized by 27 scalars in addition to the 27 charges.
One can interpret the 27 scalar parameters as
 26 scalars
plus the ADM energy. 
 Each charge corresponds to a type of
soliton or string. Thus we expect the solution to again be characterized
by the number of solitons and anti-solitons. For an extremal black hole,
the entropy can be written in the $E_6$ invariant form 
\jpup , \rk\ 
\eqn\eeninv{ S= 2\pi |T_{ABC} V^A V^B V^C|^{1/2} ~, }
where $V^A$ is the 27 dimensional charge vector and $T_{ABC}$ is a
symmetric cubic
invariant in $E_6$. For the non-extremal black holes, the above argument
suggests that one
can introduce two vectors $V_i^A \ i=1,2$ which represent the number
of solitons and anti-solitons. Although we have not done the calculation,
the general black hole entropy might take the
$E_6$ invariant form
\eqn\geninv{ S = 2\pi \sum_{i,j,k} |T_{ABC} V_i^A V_j^B V_k^C|^{1/2}~, }
where $i,j,k=1,2$, $V^A_1$ indicates the number of charges and 
$V_2^A$ the number of anticharges, each is a vector in the 27 of $E_6$.
The entropy of non-extremal
black holes can be represented in terms of charges and anti-charges
in many different (equivalent) ways which
are related by $E_6$ transformations.
Now we see that our choice of D-onebranes, D-fivebranes and momentum
was like a choice of basis and other configurations are 
related by $E_6(Z)$ U-duality transformations.

One can similarly construct rotating black holes in 
five dimensions. The spatial rotation group is 
SO(4)$_E\sim$ SU(2)$_R \times$SU(2)$_L$. We 
we can view the angular momentum as  a 
4$\times$4 antisymmetric matrix. We can choose a basis
such that it reduces to 2$\times$2 blocks, each block 
corresponds to a rotation on a plane and there are
two orthogonal planes. The angular momentum is characterized 
by the angular momentum eigenvalues $J_1,J_2$ on these two 
 planes. 
Also the angular momenta are characterized by the U(1) charges 
$F_R/2, F_L/2$ which are two eigenvalues of the SU(2)'s
(we define $F_{R,L}$ to have integer
eigenvalues).
 We have
\eqn\angfive{
J_1 = {1 \over 2 }(F_R + F_L ), ~~~~~~~~~~~J_2 =
{1 \over 2 }( F_R - F_L )~.
}
The solution with angular momentum can be found 
in 
\spn , \vbd , \cveticang , we will be just interested
in the entropy of that solution in the extremal limit,
for which the mass is the minimum
consistent with a given angular momentum and charges.
The entropy is then
\eqn\entextrotf{
S_{ext} = 2  \pi \sqrt { N Q_1 Q_5 - J_1 J_2 }~.
}
For $J_1=J_2 $ the solution is also BPS \spn .


\subsec{Black hole solutions  in four dimensions}

Now we turn to the more realistic case of four dimensional
black holes. It is still not totally realistic since the
compactification we will consider is on $T^6$ which is 
not the one that describes our four dimensional world.
The supergravity theory however contains black hole solutions
which are exactly those of General Relativity. The difference
between the two theories is that the N=8 supergravity theory
one obtains by compactifying on $T^6$ has many more 
gauge fields (28 of them) and  massless scalars (70 of them).
Black hole solutions are characterized by 
56 charges, 28 electric and 28 magnetic. 
One hopes that the general features of black hole physics
will not depend too much on the content of the theory, as
long as it includes gravity and one is studying 
black holes with   the same  metric  as the ones 
appearing in 
General Relativity. In fact some of the solutions we 
study are also solutions in general relativity.

\vskip .1in 
\noindent
{\it 2.6.1  Extremal black holes in four dimensions}
\vskip .1in

Let us start with the extremal black holes 
\ms , \cliffordfd . 
Taking the configuration of 1D-branes, 5D-branes
and momentum that we had in $d=5$ 
and putting it on $T^6$ we obtain a 
black hole solution that preserves 1/8 of the supersymmetries.
In order  to put it on $T^6$ one has to form a lattice of
the extremal five dimensional black holes \tenfived\ and 
define new harmonic functions as in \lattice . 
This makes all harmonic functions to depend on 
$1/r$ where now $r$ is the spatial distance in 
1+3 dimensions. 
The unbroken supersymmetries are given by 
\spinor . Now we do a $T$ duality transformation
(along the direction 4)  to 
the IIA theory and we get a system of 2D-branes, 6D-branes
and momentum. In addition we flip the chirality of 
the ten dimensional spinor $\epsilon_R$.
We will have $\epsilon_R^{IIA}  = \Gamma^4 \epsilon_R^{IIB} $, 
so that
the chirality that is flipped in $\epsilon_R$ is that
of the ``external" SO(4)$_E$. Of course, only the SO(3) subgroup
corresponding to spatial rotations in the 
 directions $1,2,3$ is a symmetry 
of the solution.
However, this black hole has zero area and
has a singular geometry at the horizon. The reason 
is that some of the scalar fields are unbalanced, 
for example, we can see from \pbrane\ that the dilaton
field will  not go to a constant as we approach the horizon,
$e^{-2 \phi } = f_2^{-1/2} f_6^{3/2}$.
It is interesting that one can put an additional 
type of charge without breaking any additional supersymmetry.
This charge has to be a solitonic fivebrane, it is the 
only one allowed by supersymmetry that is not just 
a U-duality tranformation of the others.
This, in a sense, is analogous to putting left moving
oscillations on a macroscopic heterotic string
\cmp\ which does not break any additional supersymmetry. 
It also has the virtue of balancing all the scalars,
for example the dilaton now behaves as $e^{- 2 \phi } =
f_2^{-1/2} f_6^{3/2} f^{-1}_{s5} $.
In order to 
be more precise let us say that our torus is
$T^6 = T^4\times S_1' \times S_1$
and we have the 6D-branes wrapping all $T^6$, the 
2D-branes wrap $ S_1' \times S_1$ (directions $4,9$),
 the solitonic
fivebranes wrap $T^4\times S_1$ (directions $5,6,7,8,9$)
and the momentum
is flowing along $S_1$ (direction $9$).
 Notice that the momentum
flows {\it parallel } to the fivebranes and the two D-branes. 

We can see from \susydbrane ,\susymom\ and   \susyfive\ 
that the fivebrane does not break any additional supersymmetry.
The
final configuration still preserves 1/8 of the original
supersymmetries. Decomposing the surviving spinor in 
terms of SO(1,1)$\times$SO(4)$_E\times$SO(4)$_I$ we find
\eqn\susyfour{
\Gamma^4 \epsilon_R = \epsilon_L = 
\epsilon_{SO(1,1)}^+ \epsilon_{SO(4)}^+ \epsilon_{SO(4)}^+~.
}

The extremal four dimensional black hole, constructed this
way, written in ten dimensional string metric, has the
form \tseytlin\
\eqn\tenfourd{
\eqalign{ d s^2_{str} = &  f^{-{1\over 2}}_2
f^{-{1\over 2}}_6 
 \left( -dt^2 + dx_9^2 +
k (dt-dx_9)^2 \right ) + 
f_{s5}  f^{-{1\over 2}}_2 f^{-{1\over 2}}_6 dx_4^2 + \cr
 + & f^{{1\over 2}}_2 f^{-{1\over 2}}_6 ( dx_5^2  + \cdots + dx^2_8 )
+
 f_{s5}  f^{{1\over 2}}_2 f^{{1\over 2}}_6 ( dx_1^2 +  \cdots + dx^2_3)~,
\cr
e^{- 2 (\phi_{10} -\phi_\infty) } =  &  f_{s5}^{-1} ~ f_2^{-{1\over 2}}~ 
f_6^{{3\over 2}} ~,
\cr
H_{ij4}   = &{1\over 2} \epsilon_{ijk} \p_k f_{s5}     ~~~~~~~~~~
i,j,k = 1,2,3~,        \cr
C_{049} = & {1 \over 2} ( f_2^{-1} -1 )~, \cr
(d A )_{ij} = & {1\over 2} \epsilon_{ijk} \p_k f_6 ~~~~~~~~~~
i,j,k = 1,2,3 ~,
}}
where $\epsilon_{ijk}$ is the flat space epsilon tensor.
The harmonic functions are
\eqn\harmonicfo{
f_2 = 1 + { c^{(4)}_2 Q_2 \over r},~~~~~
f_5 = 1 + { c^{(4)}_{s5} Q_5 \over r},~~~~~
f_6 = 1 + { c^{(4)}_6 Q_6 \over r},~~~~~
k =  { c^{(4)}_P N \over r}~,
}
where the coefficients $c^{(4)}$'s are given in \constfour\ and
the charges $Q_2,Q_5,Q_6,N$ are integers.
Calculating the entropy we find 
\eqn\entrofe{
S = { A_4 \over 4 G_N^4 } = 2 \pi \sqrt{Q_2 Q_5 Q_6 N }~,
}
which is, as \entropyext , 
 U-dual and independent of the moduli.

\vskip .1in
\noindent
{\it 2.6.2 
Non-extremal black holes in four dimensions}
\vskip .1in

In a similar way as we did for five dimensions one can 
construct the  non-extremal four dimensional solution.
After doing the dimensional reduction to four dimensions 
the Einstein metric reads \cvetd\foot{
We  use the classical solution from \cvetd\ but with our 
quantization condition for the charges derived  in
section (2.4).}
\eqn\metricfour{
\eqalign{
ds^2 &= - \chi^{-1/2}(r)\(1 - {r_0  \over r}\) dt^2 +
 {\chi^{1/2}(r) }\left[
\left( 1 -{r_0 \over r} \right)^{-1}  dr^2
+  r^2 (d\theta^2 + \sin^2\theta d\phi^2)\right]~,\cr
\chi(r) &=
\(1 +{ r_0 \sinh^2
\alpha_2  \over r}\)\(1 + { r_0  \sinh^2 \alpha_5 \over r}\)
\(1+ { r_0  \sinh^2 \alpha_6 \over r}\)\(1 + { r_0
\sinh^2 \alpha_p \over r }\)~.\cr}
}
This metric is parameterized by the
five  independent quantities $\alpha_2$, $\alpha_5$,
$\alpha_6$, $\alpha_p$ and $r_0$.
 The event horizon lies at $r=r_0$. The special case $\alpha_2 = \alpha_5
 =\alpha_6=\alpha_p$ corresponds to the Reissner-Nordstr\"om metric
\renof , so that we see that, as we promised, the 
General Relativity solution is among the cases studied.
The overall solution contains three additional parameters
 which are related
to the asymptotic values of the three scalars. From the 
ten-dimensional
viewpoint, these are the product of the radii of $T^4$, $
V = R_5 R_6 R_7 R_8$,
and the  radii of $S^1$ and $ S'^1$, 
$R_9$ and $R_4$, and they appear in the quantization condition
for the charges.
There are, in addition, U(1) gauge fields excited, corresponding
to the four physical charges. One is 
the Kaluza Klein gauge field coming from the component
$G_{09} $ of the metric, which carries the momentum charge, $N$.
Then we have a RR gauge field coming from the component
$C_{049}$ of the three form RR potential which carries the 
2D-brane charge, $Q_2$. The 6D brane charge, $Q_6$,
 appears as magnetic
charge for the one form RR potential $A_\mu$, and finally the
fivebrane charge, $Q_5$ also appears as magnetic charge for
the gauge field coming  from the NS antisymmetric tensor with
one index along the direction $\hat 4$, 
 $ B_{\mu 4} $.

The physical charges are expressed in terms of these
quantities as \hlm\
\eqn\charges{
\eqalign{
Q_2  &= {  r_0 V \over g }
\sinh 2\alpha_2 ~, \cr
Q_5 &= { r_0  R_4  }
\sinh 2\alpha_5 ~, \cr
Q_6 &= {  r_0 \over  g}
\sinh 2\alpha_6 ~, \cr
N &=  {  r_0 V R_9^2 R_4 \over g^2 }
\sinh 2\alpha_p ~, \cr}
}
where
we have again set
$\alpha'=1$ and from \gten\  
the four-dimensional Newton constant becomes
 $G^4_N =  g^2/(8 V R_4 R_9)$.

The ADM mass of the solution is
\eqn\admmass{
M= { r_0  V R_4 R_9 \over g^2}
 (\cosh 2\alpha_2+
\cosh 2\alpha_5 + \cosh 2 \alpha_6 + \cosh 2 \alpha_p )
}
and the Bekenstein-Hawking entropy is
\eqn\bhentropy{
S ={A_4\over 4G^4_N } = {8 \pi r_0^2 V R_4 R_9 \over g^2} \cosh \alpha_2
\cosh \alpha_5 \cosh \alpha_6 \cosh \alpha_p ~.
}

There are three nontrivial scalar fields present
in the solution and
associated with these scalar fields
are three pressures (scalar charges)
\eqn\scharge{
\eqalign{
P_1 &= { r_0 V  R_4 R_9 \over g^2} (\cosh 2\alpha_2 +
\cosh 2 \alpha_6 - \cosh 2\alpha_5 - \cosh 2\alpha_p) ~,\cr
P_2 &= { r_0 V  R_4 R_9 \over g^2} (\cosh 2 \alpha_2-
\cosh 2\alpha_6) ~,\cr
P_3 &= { r_0 V  R_4 R_9 \over g^2} (\cosh 2\alpha_5 -
\cosh 2\alpha_p ) ~.\cr}
}

As we did for the five dimensional black hole in
section (2.5) we calculate the
values for the mass and scalar charges
of each type of brane or string.
This can be calculated from the solution we have presented by
taking the four extremal limits: $r_0 \rightarrow 0, ~
\alpha_i \rightarrow 
\pm \infty$ with $Q_i$ and $\alpha_j \ (j\ne i)$ fixed.
We find that D-twobranes have mass and pressures \hlm\
\eqn\twobranes{
M=P_1 =P_2 = { R_4 R_9 \over g }~,~~~~~~~~
P_3 = 0 ~,}
while for the D-sixbranes we have
\eqn\sixbranes{
M = P_1 = -P_2 = { V  R_4 R_9 \over g }~,~~~~~~~~~P_3 =0~.
}
For the solitonic fivebrane we have
\eqn\fivebrane{
M=  -P_1 = P_3 = { V R_9 \over g^2 }~,~~~~~~~~~~~~P_2 =0 ~,}
and for the momentum we find
\eqn\momentum{
M= -P_1 = -P_3 = { 1 \over R_9}~, ~~~~~~~~~~~~~P_2=0~.}

Using these relations plus the charges  \charges\  we
trade in the eight parameters of the
solution for the eight quantities $(N_R, N_L,
N_2,  N_{\bar 2}, N_5,  N_{\bar 5}, N_6,  N_{\bar 6})$
which are the numbers of
right(left)-moving momentum
modes, twobranes, anti-twobranes, fivebranes,
anti-fivebranes, sixbranes and anti-sixbranes.
We do this by matching the mass \admmass ,
pressures \scharge ,
and gauge charges \charges\ with those of a collection of
noninteracting branes. This leads to
\eqn\branenum{
\eqalign{
N_R & = {r_0  V R_9^2 R_4 \over 2 g^2} e^{2\alpha_p}~,  \cr
N_2 &= {r_0  V  \over 2 g} e^{2\alpha_2}~, \cr
N_5 &= {r_0  R_4 \over 2}
e^{2\alpha_5 }~,  \cr
N_6 &= {r_0  \over 2 g} e^{2\alpha_6}~, \cr}
\qquad
\eqalign{
N_L  &= {r_0 V R_9^2 R_4 \over 2 g^2} e^{-2\alpha_p}~, \cr
 N_{\bar 2} &= {r_0  V  \over 2 g} e^{-2\alpha_2}~, \cr
 N_{\bar 5} &= {r_0 R_4 \over 2 }e^{-2\alpha_5}~, \cr
 N_{\bar 6} &= {r_0    \over 2 g} e^{-2\alpha_6}~. \cr}
}

In terms of the brane numbers, the ADM mass is
reexpressed as
\eqn\badm{
M = {1\over R_1} (N_R+N_L)+ {R_9 R_4 \over g} (N_2+
 N_{\bar 2}) + {V R_9\over g^2}(N_5+ N_{\bar 5}) +
{V R_9 R_4 \over g}(N_6+ N_{\bar 6})~,
}
the gauge charges are simply differences of the
brane numbers, and the other parameters are
\eqn\vollen{
V= \sqrt{ {N_2  N_{\bar 2} }\over {N_6  N_{\bar 6} }} ~~,
\qquad R_4 = \sqrt{ {{N_5  N_{\bar 5}}\over {g^2 N_6
 N_{\bar 6}}}}~~, \qquad R_9^2 R_4 = \sqrt{ {{g^2 N_R N_L}
\over {N_2  N_{\bar 2} }}}~.
}

The entropy \bhentropy\ then takes the surprisingly simple form
\eqn\entropyb{
S = 2\pi (\sqrt{N_R} +\sqrt{N_L})(\sqrt{N_2}+\sqrt{ N_{\bar 2} })
(\sqrt{N_5}+
\sqrt{ N_{\bar 5}})(\sqrt{N_6}+\sqrt{ N_{\bar 6} })~.
}
This is the analog of \smira\ for four-dimensional black holes. When
one term in each factor vanishes, the black hole is extremal \entrofe .
 Although we cannot derive the general
formula from counting string states, we will do so in certain limits
corresponding to near-extremal black holes.

Since the full theory should  be $E_7$ invariant we should
be able to write the general entropy formula in an invariant way.
If we denote by $V^A_1$ the 56-dimensional vector giving the number of
solitons and by
$ V^A_2 $ the number of anti-solitons, the formula for the
entropy  takes the form \kara \koll\
\eqn\eseven{
S = 2 \pi \sum_{i,j,k,l} \sqrt{ T_{ABCD} V^A_i V^B_j V^C_k V^D_l }~,
}
where $T_{ABCD}$ is the $E_7$ quartic invariant.

We now consider adding rotation to the black holes discussed above.
Since the rotation dependent terms in the solution fall off faster
at infinity than the charges, the definition of the brane numbers
\branenum\ is unchanged.
If we  take nearly extremal black holes with
$ N_{\bar 2} \sim  N_{\bar 5} \sim   N_{\bar 6} \sim 0$,
 and $R_1$ large,
the Bekenstein-Hawking entropy  takes the form \cveticang \
\eqn\neent{
S=2\pi  \( \sqrt{ N_R N_2 N_5 N_6} +
\sqrt{ N_L N_2 N_5 N_6 - J^2 } \)~.
}
where $J$ is the angular momentum of the black hole.

\vfill\eject

\newsec{\bf D-BRANE DESCRIPTION OF BLACK HOLES}

\vskip .2truein

In this chapter we will describe some 
properties of the black holes 
that we studied  in chap. 2 in terms of
the D-brane string solitons described
in chap. 1. 
The black holes we study have non-zero horizon
area and contain, as special cases, the charged black
holes of general relativity.
The description that
will emerge depends  on some detailed properties of
the string theory D-brane solitons which 
are obvious from just 
the supergravity theory.
The key property that will be used is that when many
D-branes sit at the same point there is a large number
of massless states coming from open strings with 
ends attached to different branes \polchinskinotes .

We will start with the five dimensional black hole which
is  simpler because it   involves  only three
different kinds of charge. Then we treat  the
four dimensional case. In both cases
we will start with the BPS extremal black holes 
where the  calculation of the entropy 
 can be justified on the
basis of supersymmetry and then we will 
explore the near-extremal limits. 

\newsec{Extremal Five dimensional black holes.}

We consider the type IIB theory on $T^5 = T^4\times S^1$.
We consider a configuration of
$Q_5$ D-fivebranes wrapping the whole $T^5$,
$Q_1$ D-strings wrapping the $S^1$ and momentum $N/R_9$ 
along the $S^1$, choosing this $S^1$ to be in the
direction $\hat9$. All charges $N,Q_1,Q_5$ are integers \einstein . 

Since extremal D-branes are boost invariant along the
directions parallel to the branes they cannot carry
momentum along $S_1$ by just moving rigidly.
Our first task will be to identify the D-brane excitations
that carry the momentum. 
In the discussion of the oscillating fundamental string in 
sec. (2.2) the 
momentum was carried by the oscillations.
However, just oscillations of the branes do not have
enough entropy to match the classical result.
As we saw in the D-brane section, oscillations of the 
branes are described by massless open strings with both ends 
attached to 
the same  brane.
There are many types of open strings to consider: those that
go from one 1-brane to another 1-brane, which  we denote  as (1,1) strings,
as well as the corresponding  (5,5), (1,5) and (5,1) strings (the last two
being different because the strings are oriented). 
We want to excite these strings and put some momentum on them.
As it was shown for  oscillating D-branes in 
section 1.3  exciting some of them makes others massive so we
have to see what is the way to excite the stings so that
a maximum number remains massless, since  this  configuration will
 have the 
highest entropy. 
Let us work out the properties of (1,5) and (5,1) strings.
The string is described by the action \actionstr\ where 
two of the coordinates have Neumann-Neumann boundary conditions 
($X^0,X^9$), four coordinates have Dirichlet-Dirichlet boundary
conditions ($X^1,X^2,X^3,X^4$) and the other four have
Neumann-Dirichlet conditions ($X^5,X^6,X^7,X^8$).
The vacuum energy of the worldsheet bosons is 
$ E = 4(-1/24 + 1/48)$. Consider the NS sector for the
worldsheet fermions, 
the  4 that
are in the ND directions will end up having R-type quantization
conditions. The net fermionic vacuum energy is $ E= 4( 1/24 -  1/48) $
and exactly cancels the bosonic one.
This vacuum is a spinor under SO(4)$_I$,  is acted on by
$\Gamma^5,\Gamma^6,\Gamma^7,\Gamma^8$,
 and obeys the GSO chirality condition
$ \Gamma^5\Gamma^6\Gamma^7\Gamma^8 \chi = \chi$. What remains is a
two dimensional representation. There are
two possible orientations and they can be attached to
any of the different branes of each type. This gives a total
of $ 4 Q_1 Q_5$ different possible states for these strings.
Now consider the Ramond sector, the four internal fermions
transverse to the string will have NS type boundary conditions. The
vacuum again has zero energy and is an SO(1,5) spinor and a 
spacetime  fermion.
Again the GSO condition implies that only the positive
chirality representation of SO(1,5) survives.
When it is also  left moving  only the $ 2^+_+$ under
SO(1,1)$\times$SO(4)$_E$ survives. This gives the same number of states as
for the bosons.
Note that the fermionic (1,5) (or (5,1)) strings carry angular
momentum under the spatial rotation group SO(4)$_E$ but the 
bosonic (1,5) (or (5,1)) strings do not carry angular momentum.

\vskip 1cm
\vbox{
{\centerline{\epsfxsize=4in \epsfbox{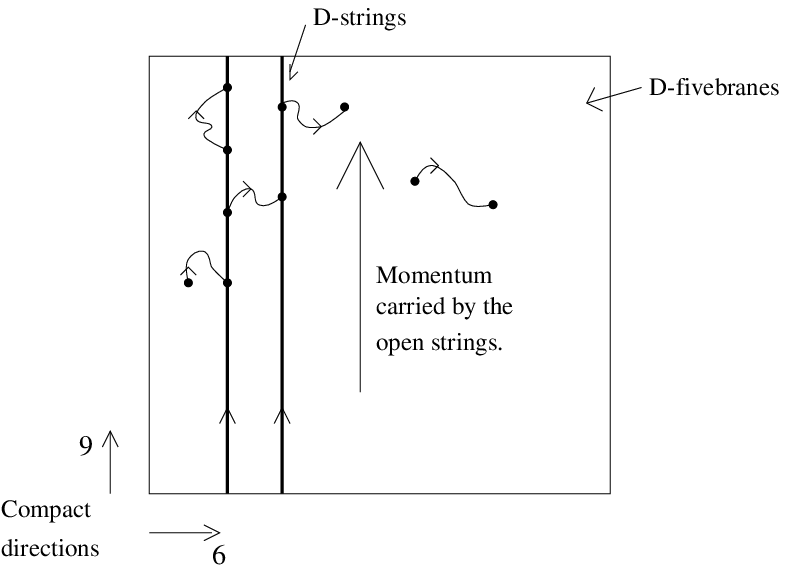}}}
{\centerline{\tenrm FIGURE 5:
Configuration of intersecting D-branes. We show  two of
the internal
}}
{\centerline{
dimensions and several types of open strings. The open
strings going between
}}
{\centerline{
1 and 5 branes are the most relevant for the black holes
that we analyze.
}}
}
\vskip .5cm

In order to be more definite about the number of massless
degrees of freedom 
it is necessary to know something about the interactions of
these open strings. It was shown in 
\dgl\ that this interaction Lagrangian is determined by 
supersymmetry and  gauge symmetry.
 The 1+1 dimensional field
theory is a (4,4)  supersymmetric theory, which has  the same
amount of supersymmetry as $N=2$ in four dimensions or $N=1$ in 
six dimensions.
For simplicity we will discuss just the bosonic part of the
Lagrangian. We saw in chapter 1 that the Lagrangian for the
(1,1) or (5,5) strings is the dimensional reduction 
of the $d=10$ Yang Mills Lagrangian and  a way to see this
was to do T-duality transformations so that the 
$p$-branes became 9-branes.
In the same spirit we do T-duality transformations that
map our 1D and 5D branes into 5D and 9D branes.
Now we have a six dimensional theory on the fivebrane. 
Let us denote by $\alpha, \beta =0,1,2,3,4,9 $ the indices
along the fivebrane, by $I,J=5,..8 $ the indices perpedicular
to the fivebranes and by $\mu,\nu = 0,..,9$ the full ten
dimensional indices.
For simplicity we will concentrate only on the bosonic
part of the Lagrangian, and we will give only the
bosonic part of the supermultiplets.
We have N=1 supersymmetry in six dimensions. There are two 
possible supermultiplets, the vector multiplet and the
hypermultiplet. The vector contains a six dimensional
vector field $A_\alpha $ and its spin 1/2 superpartner.
The hypermultiplet contains four scalar fields,
and the spin 1/2 superpartners.
The ten dimensional vector $A_\mu$ describing (5,5) strings 
decomposes into 
a six dimensional vector $A_\alpha$ plus a hypermultiplet
containing the four scalars $A_I$ representing the transverse 
motion
of the fivebranes. Both of these fields are in the 
adjoint of the U($Q_1$) gauge group\foot{
The index of $Q_1$ reminds us that in the aplication of interest
these fivebranes are D-strings,  hopefully   this will  not
cause confusion.}.
The YM Lagrangian \tendym\ has N=2 supersymmetry in six dimensions
and it can, of course, be thought of as an N=1 Lagrangian.
The  (5,9) and (9,5) strings together form a hypermultiplet
with fields which transform as the product of
the fundamental of U($Q_1$) and the anti-fundamental of U($Q_5$) and
their complex conjugate. These are the fields $\chi_a^B$ discussed
above, $\chi$ was a $2^+$ spinor under the SO(4)$_I$ and 
$a$ and $B$ represent $U(Q_1)\times \bar U(Q_5) $ Chan Paton 
factors. The other two components of this hypermultiplet 
are their complex conjugate  $ \chi^{\dagger \, a }_B$.
The interaction Lagrangian is determined largely by
supersymmetry. The only allowed coupling between vector and 
hypermultiplets is the gauge coupling. 
The hypermultiplets could, in principle, have a metric
in the kinetic term but to lowest order in the
string coupling this metric is flat. 
Supersymmetry requires, however, some potential for the
hypermultiplets, this is given by the so called ``D-terms''
(no relation to ``D''-branes). They arise when we have
gauge symmetries, there are three D-terms for
 each gauge generator. One way to think about them
is through the SO(4)$_I$ language as self dual 
antisymmetric tensors $ D_{IJ} =
 {1 \over 2 } \epsilon_{IJKL} D_{KL} $, such a tensor
has three independent components. The potential is
given by $V \sim \sum_a D^{a2}_{12} + 
 D^{a2}_{13} + D^{a2}_{14} $.
D-terms have the structure $D^a \sim \phi^\dagger T^a \phi'$ 
where $\phi $ are components of the hypermultiplets and
$T^a$ are the gauge group generators. 
As an example of D-terms let us consider the pure
YM Lagrangian dimensionally reduced to six dimensions.
The potential term for the hypermultiplets comes simply 
form the commutator terms in the YM Lagrangian
\eqn\vym{
V = Tr F_{IJ} F^{IJ}  = \sum_{IJ} Tr [A_I,A_J]^2 =
\sum_a D^{a2}_{12} +
 D^{a2}_{13} + D^{a2}_{14}
}
where $D_{IJ}^a$ is defined through
\eqn\dym{
 D_{IJ}^a T^a = [ A_I, A_J ] +  {1 \over 2 } \epsilon_{IJKL}
[A_K , A_L ] 
}
In checking \vym\ with \dym\ one should use the Jacobi 
identity for the commutators. We would have obtained the same
result if we had taken the antiself dual part of
$ [A_I , A_J ]$ in \dym , this a reflection that the YM
Lagrangian \tendym\ really has N=2 supersymmetry as a six dimensional
theory. 

Now let us consider the D-terms for the full theory, they 
have the form
\eqn\dfull{
D^a_{IJ} =
f^a_{bc} (A_I^b A_J^c + {1 \over 2 } \epsilon_{IJKL} 
A_K^b A_L^c ) + \chi^\dagger T^a \Gamma_{IJ} \chi 
}
$$
 V = \sum_{ a I J } D^{a2}_{IJ}
 $$
where the index $a$ runs over the gauge group generators,
first of U($Q_1$) and then of U($Q_5$). Note that the first
term involves (5,5) or (9,9) fields depending on which
generator we consider. 
The second
term  is automatically self dual due to the chirality
condition of the spinor $\chi$. We have not been very careful
with the precise numerical normalization of these two terms
because we will not need it in what follows.
The full action has the form, up to numerical normalizations,
\eqn\fullac{\eqalign{
S = & {1 \over g} \int Tr( F_{\alpha \beta} F^{\alpha \beta})
+  Tr( F'_{\alpha \beta} F'^{\alpha \beta}) 
+ Tr [ ( \p_\alpha A_I + [A_\alpha, A_I] )^2 ] 
+\cr
+& Tr [ ( \p_\alpha A'_I + [A'_\alpha, A'_I] )^2 ] +
 |( \p_\alpha + A_\alpha^a T^a + A'^a_\alpha T^a ) \chi |^2 +
\sum_{a IJ} D^{a 2}_{IJ}~,
}}
where we denoted by $A_\alpha , A'_\alpha $ the gauge fields of
U($Q_1$) and U($Q_5$) respectively. The index $a$ in the D-terms 
runs over both gauge groups.

Now that we understand better the Lagrangian let us do 
a T-duality transformation 
back to the 1D and 5D branes. Now we have an N=4 theory
in two dimensions, the supermultiplets are just the 
dimensional reduction of the six dimensional ones and
therefore
recieve the same name.
Some of the components of the six dimensional vector
multiplet become scalars and they represent the motion 
of the D-string or the D-fivebrane transverse to the
fivebrane. The motion of the D-string on the fivebrane, 
as well as the fivebrane gauge fields transverse to the string
are hypermultiplets in this language. 

The BPS states that we are considering have only left moving
excitations. One can view these states classically as traveling
waves 
propagating along $S_1$. In order to have traveling wave solutions
the mass terms have to vanish exactly. 
If we set all the fields in the Lagrangian  to zero then 
we can have traveling waves for any field. However once we
have a wave for one field the potential terms in \fullac\
in the case of hypermultiplets or gauge couplings 
in the case of vector multiplets imply that there will
be mass terms for other waves as in \offdia . 
One should find which is the way to excite the system 
in such a way that a maximum number of particles remains 
massless. 
For determining the mass terms, as in \offdia , it is the same
to consider configurations that are $u=x^9 -t$ dependent or
not. The problem is analogous to finding the  sector of the
moduli space of vacua with the largest number of massless 
particles. 
If we give some expectation value to the scalars coming
from the six dimensional vector fields, then we see that
we are effectively   separating the strings and fivebranes
and we expect a small number of massless particles (proportional
to $Q_1 + Q_5$). Indeed there are mass terms for the
hypermultiplets in the fundamental due to the gauge 
couplings $ \chi^\dagger A_\alpha A^\alpha \chi $ \fullac . 
This mass term also implies that if the (1,5) strings condense
then there is a mass term for the transverse motion. 
In fact a configuration with a large number of massless 
particles is achieved by exciting the hypermultiplets, 
both the ones in the fundamental and the ones in the 
adjoint. This gives mass to the scalars describing the
transverse motion of the branes which means that
we have a bound state.
The total number of components of the hypermultiplets is
$4 Q_1^2 + 4 Q_5^2  + 4 Q_1 Q_5 $. In order to preserve supersymmetry
we must set the potential to zero, which  also minimizes the energy.
This implies that the D-terms \dfull\ should vanish, 
imposing $ 3 Q_1^2 + 3Q_5^2 $ constraints. In addition we should
identify  gauge equivalent configurations. The number
of possible gauge transformations is $Q_1^2 + Q_5^2 $. This 
implies that the remaining  number of  massless
degrees of freedom is $ 4 Q_1 Q_5 $. 
The counting, as we have done it here, is correct for large 
charges up to possible subleading corrections. 
One can think that what we did was to determine the dimension
of the classical moduli space of vacua of this theory and then
considered oscillations around a given vacuum. 

Let us remark for later use that in this picture the momentum 
is carried by the hypermultiplets. The bosonic components
of the hypermultiplets do not carry angular momentum under SO(4)$_E$
of the
external rotations, while the fermions do indeed carry it. 
In fact the SO(4)$_E \sim $SU(2)$_L \times $SU(2)$_R$ symmetry
appears as an R-symmetry of the theory and  left or
right moving fermions carry spin under SU(2)$_{L,R}$ respectively \spn .

The state 
counting is the same as that of the left moving oscillator modes
of $ 4 Q_1 Q_5 $   superconformal fields.
These modes will be carrying purely left moving momentum. 
In order to calculate the entropy we notice that we
have a gas of left moving particles  with
$ N_{B,F} = 4 Q_1 Q_5 $
 bosonic and fermionic species with  energy 
$E=N/R_9$
on a compact one dimensional space of length $L= 2 \pi R_9 $.
The standard entropy formula gives \sv \cama\
\eqn\entropyd{ S_e = \sqrt{ \pi (2  N_B + N_F) E L/6 }   
=2 \pi\sqrt{ Q_1Q_5 N}~, }
in perfect agreement, including the numerical coefficient, 
 with \entropyext .

It is interesting to understand the relation of this
description in terms of open strings and the 
approach taken in the original derivation \sv .
In \sv\ the D-strings were viewed as intantons on
the U($Q_5$) fivebrane gauge theory \dgl \vafacount .
 Due to some
Chern Simons couplings these instantons carry RR
1D-brane charge \dgl . The moduli space of these instantons
is given by a (4,4) superconformal field theory with 
central charge $c = 6 Q_1 Q_5$. The $Q_5$ factor comes 
basically from all the possible orientations
of the instanton, and this is where the 
entropy comes from (for large $Q_5$). These instantons
live on the fivebrane and it is only when they 
shrink to zero size that they become a D-string and
are allowed to leave the fivebrane. 
Note that the steps followed in determining the number
of massless open string modes above is very similar
to calculations of instanton moduli spaces 
in \wittensmall . 

In our previous argument we implicitly  took
the D-strings and the 
fivebranes to be singly wound.
For large $N$, $N \gg Q_1 Q_5 $,  the entropy \entropyd\ is the
same no matter  how we  the branes are wound. 
However for $ N \sim Q_1 $ the winding starts to matter.
The reason is that in order for the asymptotic 
formula to be correct for low $N$ we need to have
enough states with small energies \sm . 
Let us study the effect of different wrappings.
We begin with an
analogy from
elementary quantum mechanics. Consider a circular loop of wire of unit
radius
whose center is at the origin of the $r,\theta$ plane. A bead of unit mass
moves
on the wire and for obvious reasons the angular momentum of the bead is
quantized
in integer multiples of
$\hbar$. The energy spectrum is given by
\eqn\qm{
E={l^2\over 2}
}
for all integer $l$.
 Now consider a wire which is wrapped $n$ times around the same circle.
Eq.
\qm\
 still gives the energy levels if we allow $l$ to be an integer multiple
of
$1/n$. The system simulates fractional angular momentum. The real physical
system
of wire plus bead must, of course, have integer angular momentum but the
energy
spectrum may be expressed in terms of a
``psuedo-angular momentum" which is
not
the true generator of spatial rotations but rather the generator of
rotations
of
the bead relative to the physical wire.

Next let us consider a set of $Q_1$  1-branes wrapped on $S^1$, ignoring for
the
time being, the 5-branes. We may distinguish the
various ways
the branes interconnect. For example,
 they may connect up so as to form one
long
brane of total length $R'= R Q_1$. At the opposite extreme they might
form $Q_1$ disconnected loops. The spectra of open strings is different in
each
case. For the latter case the open strings behave like $Q_1$ species of
1 dimensional particles, each  with energy spectrum given by integer
multiples of
$1/R$. In the former case they behave more like a single species of 1
dimensional
particle living on a space of length $Q_1 R$. The result  \dasmathur\
 is a
 spectrum
of
single
particle energies given by integer multiples of $1\over{Q_1 R}$  .
 In other
words
the system simulates a spectrum of   fractional charges. For consistency the
total
charge must add up to an integer multiple of $1/R$ but it can do so by
adding
up
fractional charges. Note that in this case, as opposed to the
bead and wire example, the branes by themselves cannot carry any
momentum since they are invariant under boosts along directions
parallel to the branes.

Now let us return to the case of both 1 and 5 branes. By suppressing
reference
to
the four compact directions orthogonal to $x^9$ we may think of
the 5 branes
as
another kind of 1 brane wrapped on $S^1$. The 5-branes
 may also be connected
to
form a single multiply wound brane or several singly wound branes. Let us
consider the spectrum
 of (1,5) type strings (strings which connect a 1-brane
to
a five-brane) when both the 1 and 5 branes each
form a single long brane. The 1-brane has total length $Q_1 R$ and the
5-brane
has length $Q_5 R$. A given open string can be indexed by a pair of indices
$[i,\bar j]$ 
labelling which loop of 1-brane and 5-brane it ends on. As a
simple
example choose $Q_1=2$ and $Q_5=3$. Now start with the $[1,1]$ string which
connects the first loop of 1-brane to the first loop  of 5-brane. Let us
transport this string around the $S^1$. When it comes back to the starting
point
it is a $[2,2]$ string. Transport it again and it becomes a $[1,3]$
 string.
It
must be cycled 6 times before returning to the $[1,1]$ configuration. It
follows
that such a string has a spectrum of a single species living on a circle of
size
$6R$. More generally, if $Q_1$ and $Q_5$ are relatively prime the system
simulates a single species
 on a circle of size $Q_1 Q_5 R$. If the $Q's$ are
not
relatively prime the situation is slightly
more complicated but the result is the same.  For example  suppose
$Q_1 = Q_5 =Q$,  again  assume the 5 and 1-branes each form a single long
brane,
then a string will return to its original configuration after
cycling
around $Q$ times. This time the system simulates $Q$ species living
on
a
circle of length $Q$. But  it is also possible to remove one loop from
either
the 1 or 5 brane and allow it to form a separate disconnected  loop. In this
case we have a system consisting of a brane of length $QR$, one of length
$(Q-1)R$ and a short loop of length $R$. Since $Q$ and $Q-1$ are relatively
prime the open strings which connect them live on an effective brane of
length
$Q(Q-1)R$. Thus there is always a way of hooking up branes so that the
effective length is of order $Q_1 Q_5 R$. In fact we will argue
that this type of configurations give the largest entropy, and will
therefore be dominant \sm .

It can also be seen from the original derivation
of the black hole entropy by
Vafa and Strominger \sv , that the system should have low
energy modes with energy of order $ 1/R Q_1 Q_5 $.
In this derivation  the degrees of freedom that
carry  the momentum were described by a superconformal
field theory on the orbifold $(T^4)^{Q_1 Q_5} /S(Q_1 Q_5)$.
A careful analysis of this theory shows that  low
energy modes are present.
For example, excitations with  angular momentum
are associated to energies $ \delta E \sim J^2/R Q_1 Q_5$ \spn 
\vbd\ so that
for small angular momentum we are having  a gap of the correct 
magnitude.

We can easily see that this way of wrapping 
the branes  gives the correct value for the extremal
entropy. Let us consider the case where
$Q_1$ and
$Q_5$ are  relatively prime. As in \cama\ the open strings have 4 bosonic
and 4 fermionic degrees of freedom and carry total
momentum $N/R$. This time the quantization length is
$R'=Q_1Q_5R$ and the momentum is quantized in
units of $(Q_1 Q_5 R)^{-1}$. Thus instead of being at level
$N$ the system is at level $N'=NQ_1Q_5$. In place of the original
$Q_1Q_5$
species we now have a single species. 
The result is
\eqn\entrocorr{
S=2\pi\sqrt{N'}=2\pi\sqrt{NQ_1Q_5}
}
 The picture is  very reminiscent of
that proposed in \larsen\
although the details differ.

The case we have been considering so far corresponds to 
black holes in $N=8$ supergravity. With these methods 
we can also calculate the entropy for black holes
in $N=4$ supergravity. This theory is the low energy 
limit of the heterotic string compactified on 
$T^5$.
These are the dyonic black holes considered by
various authors 
\larsen , \tseytlin , \tseylinother , \cvyo .
In that case, the analogous D-brane description takes place
in the type I theory, which is S-dual to the heterotic theory.
Type I string  theory indeed 
contains D-strings and D-fivebranes (but no other D-branes). 
These two D-branes correspond to the fundamental string and 
the solitonic fivebrane on the heterotic side \wittenpol .
 The D-brane counting can also
be done, and it is interesting to notice that to get the
correct result one must include the 5D-brane SU(2) degrees
of freedom found in \wittensmall .

Something remarkable has happened here.
We started with some configuration of D-branes sitting at $r=0$,
a point in 5-dimensional space. To start with, this is a
``point with nothing inside it.'' However, having put all these
open strings on the branes   we find that the configuration
matches  a solution of  the classical low energy action
such that  $r=0$ is really  a 3-sphere with non-zero
area! What happened? Well, the ten dimensional classical
solutions for D-branes show that as we get closer to the D-brane the
transverse space expands and the longitudinal space shrinks.
This configuration has expanded the transverse space in such
great way that what previously was a point is now a sphere.
The most exciting aspect of this is that the classical solution
continues beyond the horizon, into the black hole singularity, whereas,
according to the D-brane picture space simply stops at $r=0$.
States inside the horizon would have to be described by the massless
modes on the D-brane. The basic horizon degrees of freedom are
denumerated by three integers: the momentum, the index labeling the
1-brane and the index labeling the 5-brane.
When a string ``falls'' into the black hole  and crosses the
horizon, it turns into   open strings traveling on the
D-branes (see figure 2).
There should be a mapping between the closed string degrees of freedom,
like the angle on  $S_3$ where the infalling particle hit the horizon
and the open string states.
Of course, the transformation of ``physical'' space coordinates
and open strings on the D-brane could be very complicated.
All of this is reminiscent  of the ``holographic'' principle \hologram,
as well as the membrane paradigm 
\lenyspeculations ,\membrane ,
in that dynamics occurring inside the black hole would be described
as occurring on the horizon.

\vskip 1cm
\vbox{
{\centerline{\epsfysize=2in \epsfbox{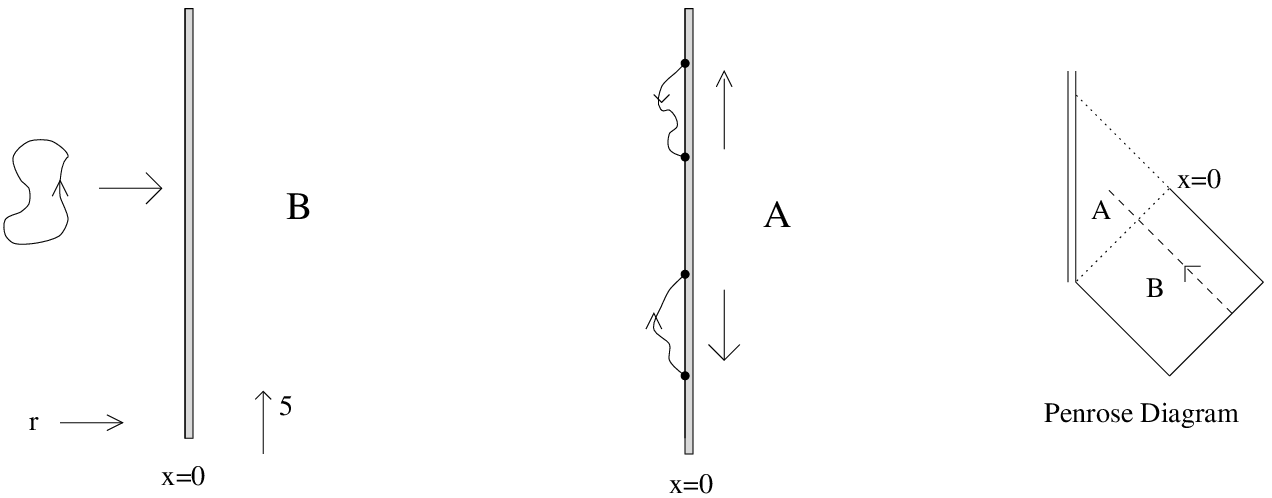}}}
{\centerline{\tenrm
FIGURE 6: D-brane description of a  string before and after
falling  through the horizon.
}}
}
\vskip .5cm

\subsec{Near-Extremal $5\,d$ Black Holes  and Hawking Radiation.}

We now turn to a discussion of nearly  extremal five-dimensional
black holes \solnfd .
Since the coupling constant \dil\ in these solutions is bounded  in space
we can choose it to be weak {\it everywhere}. 
This should be a favorable case
for examining the non-BPS states of the D-brane, doing perturbative
computations of their
interactions and comparing to the canonical expectations for the 
non-extremal
black holes \solnfd .
 We will see that for the near-extremal case the
 agreement between the two approaches
is just as impressive as in the extremal case. There is, however, a hitch:
the presence of a large number of D-branes ($Q_1, Q_5 \gg 1$) amplifies the
effective open string coupling constant and, in principle, renders any
perturbative analysis of horizon dynamics unreliable
\sv ,\hstro ,\polchinskinotes . We think the situation may not be so
desperate and will present  a (non-rigorous!) argument that
open string loop corrections might not, 
in fact, change the essential physics.
The near-extremal entropy was also calculated for D-branes that 
do not excite the dilaton, for example the three brane, but
the factors do not quite agree \gkp \klts . In those cases the
corresponding black holes have scalar fields blowing up as we 
approach the horizon. For the cases we consider the 
agreement is precise. It will be interesting to 
understand the origin of the numerical 
disagreement for the calculations
in \gkp \klts .

We perturb away from the BPS limit in a macroscopically small but
microscopically large fashion ($ M \gg \delta M \gg $ mass of typical
excitations).  There are many ways to do this by adding stringy
excitations to the basic D-brane configuration. We are interested in those
excitations which cause the entropy to increase most rapidly with added energy.
One could add fundamental string modes traveling on the torus, but they have
too small a central charge to be relevant. Massive open or closed string
excitations also give a subleading contribution since the entropy of a gas
of these excitations increases at most as $\delta M^{2/3}$, and we will
find a leading  contribution going as $\delta M^{1/2}$.
One could have five brane excitations in any direction, but that entropy
increases as $\delta M ^{p \over p+1 } $ for a membrane
of $p$ dimensions. So  we conclude  that the most
important will be the ones along the string.
There are various  modes associated to the open strings going between
the various branes. Some of these are massive due to the presence
of a large number of left movers. The ones that are massless will
give the dominant contribution and correspond, as in the case of
left movers, to  oscillations on the moduli space of vacua. 
We are saying that  the first right moving strings to be 
excited will be the ones that continue keeping the 
D-terms \dfull\ zero. 
Note the very important fact that if the branes are multiply wrapped,
as we argued they had to be, these excitations will be very 
light, with masses of order $ 1/RQ_1Q_5 $. This alone will favor these
excitations over the ones we listed above. 
If we perturb away from a purely left-moving extremal background by adding
$\delta N_R$ right-moving oscillations, we also have to add
$\delta N_L=\delta N_R $ left-moving oscillations to keep the total
$N= N_L-N_R$ charge fixed. These oscillations have 4 bosons and 4 fermions,
so the central charge is the  same as it was before. The change in
left-moving entropy is proportional to $\sqrt{ N + \delta N_R }- \sqrt N $
and is of order $\delta N_R/N_L$. The change in right-moving entropy is, 
however,
of order $\sqrt{\delta N_R/N_L}$ and dominates.
More specifically, we find that
\eqn\entroright{
\left. { \Delta S \over S_e}\right|_{oscill}
= \sqrt{ \delta N_R \over N} ~.
}
This result agrees with \smira\ when the number of antibranes
is zero, $N_{\bar 1}= N_{\bar 5}=0$. We can make these numbers
to be zero by taking $R_9$ to be very large \dbranes . In that case we have
the long string limit considered in \hstro . 
Notice that we are using here the number of branes and antibranes
obtained in \dbranes .

If  we want  
to consider
more general near-extremal black holes,
 we need to find the contribution to the entropy due to the 
addition of a small amount of antibranes. We have already calculated
the increase in entropy due to a small amount of rightmovers 
\entroright\ we need to find the corresponding increments
due to $\delta N_{\bar 1}, \delta N_{\bar 5}$.
They are presumably independent
and should be added to get the total entropy increment. We have already
calculated the increase due to the right and left movers, but it is not
obvious what the entropy increase due to the anti-branes will be. There is
however a U-duality transformation  
\hull ,\sentduality ,\vafacount ,\vafaintersecting\
that, for example, turns anti-1-branes into right moving momentum states
at the price of some transformation of coupling and compactification radii.
Since the entropy increase is independent of the coupling constant and the
compactification radii, we will take the duality argument as telling us that
the counting of the brane-antibrane states is just the same as the counting
of the left- and right-moving oscillator states. The net result for the
entropy increment is
\eqn\entroantione{
\left. { \Delta S \over S} \right|_{anti-1-branes}
=  \sqrt{ \delta N_{\bar 1} \over Q_1 }.
}
Since the same argument applies to $\delta N_{\bar 5}$.
These increments were calculated explicitly in limit in which
they are dominant, for small $R_9$ or small $V$, in \hms .
The nontrivial, not completely justified,
 assumption is that we can extrapolate those
results to a regime there the three contributions are comparable.
Making this naive extrapolation we find that
the total entropy of the non-extremal solution is
\eqn\entropynon{  \left. {\Delta S \over S}\right|_{total}
= \sqrt{ \delta N_R \over N} +
\sqrt{ \delta N_{\bar 1} \over Q_1 } + \sqrt{ \delta N_{\bar 5} \over Q_5 }
}
which agrees with the classical formula \smira\ when the antibrane
numbers are all small. 
In the \RN\ case we can see from \balance\ and \dbranes\ that
the the three terms in \entropynon\ are equal and  in 
terms of the mass above extremality we get
\eqn\entroreno{
\left. { \delta S_e \over S}\right|_{string} = { 3 \over \sqrt{ 2 } }
\sqrt{\delta M \over M_e}
}
This is the standard Bekenstein-Hawking result for the strict five-dimensional
\RN\ black hole, with the correct normalization and functional dependence
on mass. Although the arguments that led us here are less than rigorous
(especially the duality argument for entropies associated with branes and
antibranes), the simple end result gives us some confidence in the
 intermediate
steps.

These non-BPS states will decay. The simplest decay process is
a collision of a right moving string excitation with a
left moving one to give a closed string that leaves the
brane. We will calculate the  emission rate for
 chargeless particles, so that the basic process is a
 right moving open string with momentum $p_9 = n/R_9Q_1 Q_5$ colliding
with a left moving one of momentum $p_9 = - n/R_9Q_1 Q_5$ to give a closed
string of energy $k_0 = 2 n/RQ_1 Q_5 $.
Notice that we are considering the branes to be multiply wound
since that is the configuration that had the highest entropy. 
If the momenta are not  exactly
opposite the outgoing string  carries some momentum in the
$9^{th}$ direction and  we get  a charged
particle. Notice that the momentum in the $9^{th}$ direction of
the outgoing particle has to be quantized in units of $1/R_9$. 
This means that outgoing charged particles have a very large mass,
and we see that they will be thermally suppressed. All charged 
particles will have a masses
 of at least the compactification scale.
\vskip 1cm
\vbox{
{\centerline{\epsfxsize=4in \epsfbox{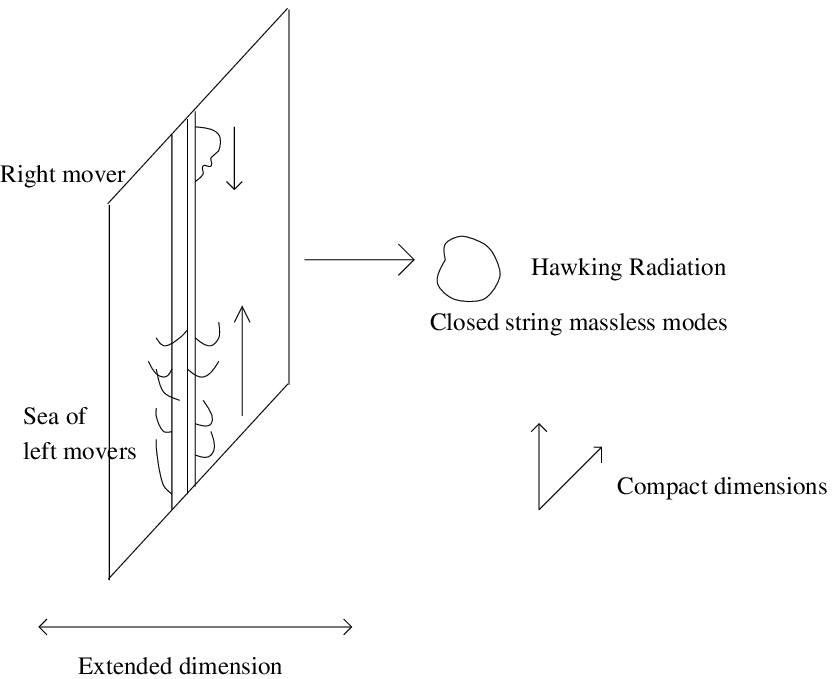}}}
{\centerline{\tenrm FIGURE 7:
D-brane picture of the Hawking radiation emision process.
}}
}
\vskip .5cm

We will calculate the rate for this process according to
the usual rules of relativistic quantum mechanics and show
that the radiation has a thermal spectrum if we do not
know the initial microscopic state of the black hole.

The state of the black hole  is specified by giving the
left and right moving occupation numbers of each of the
 bosonic and fermionic oscillators.
In fact, the nearly extremal black holes live in
a subsector of the total Hilbert space that is
isomorphic to  the Hilbert space of a one dimensional
gas of massless particles of $ 4 $ different types,
either  bosons or  fermions.
This state $|\Psi_i\rangle$ can then emit  a
closed string and become $|\Psi_f\rangle$.
The rate, averaged over initial states and summed over final states,
as one would do for calculating the decay rate of an unpolarized
atom, is
\eqn\rateunpol{
d \Gamma \sim  { d^4  k \over k_0 } { 1 \over p_0^R p_0^L V R }
\delta( k_0 -( p_0^R + p_0^L) ) \sum_{i,f}
\left| \langle \Psi_f| H_{int} | \Psi_i \rangle  \right|^2
}
We have included the factor due to 
the compactified volume  $R V$.  
The relevant string amplitude for this process is given
by a correlation function on the disc with two insertions
on the boundary, corresponding to the two open string
states and an insertion in the interior, corresponding to
the closed string state (see figure 3).
 The boundary vertex operators
change boundary conditions for four of the coordinates when 
we are dealing with a (1,5) or (5,1) string.
We consider the case when the outgoing closed string is
a spin zero  boson in five dimensions,
so that it corresponds to the dilaton,  the
internal metric,  internal $B_{\mu\nu}$ fields,
or internal components of RR gauge fields.
This disc  amplitude, call it  $\cal A $,  is proportional to the
string coupling constant $g$
and to $k_0^2$ \hk . The reason for this
last fact is that it has to vanish when we go to zero
momentum, otherwise it would indicate that there is
a mass term for the open strings (since one can vary the vacuum
expectation value of the corresponding closed string fields
continuosly). It cannot be linear either  since
the
amplitude is symmetric   under
$ k_0 \ra -k_0$ and $X_0 \ra - X_0 $, at least  for these polarizations of
the vertex operators. 
In conclusion, up to numerical factors,
\eqn\stringamplitude{
 { \cal A} \sim g k_0^2.
}

Note that performing the average over initial and sum over final states
will just produce a factor of the form $ \rho_L(n) \rho_R (n)$ with
\eqn\rhoright{
\rho_R (n) = {1 \over N_i} \sum_{i} \langle \Psi_i | a^{R \dagger}_n
 a^R_n | \Psi_i
\rangle
}
where $N_i$ is the total number of initial states and
$ a_n^R $ is the creation operator for one of the $ 4  $
bosonic open
string states. The factor $ \rho_L(n) $ is similar.
Since we are  just
averaging
 over all possible initial states with given value of $\delta N_R$,
this corresponds to taking the expectation value of $a^\dagger_n a_n$ in
the microcanonical ensemble with total energy $E_R = \delta N_R/R_9 =
\delta N'_R / R_9 Q_1 Q_5 $ 
of a one
dimensional gas. Because  $\delta N'_R  $ is large compared to one (but still
much smaller than $N'_L$), we can calculate \rhoright\ in
the canonical ensemble. Writing the partition function as
$$
Z = \sum_{N'} q^{N'} d(N') = \sum_{N'}  q^{N'} e^{ 2 \pi \sqrt{ N'} },
$$
doing a saddle point evaluation and then determining $q$ from
$$
\delta N_R Q_1 Q_5 = \delta N'_R = q {\p \over \p q } \log Z,
$$
we find $ \log q = - \pi \sqrt{1 / Q_1 Q_5 \delta N_R}  $.
Then we can calculate the occupation number of that level as
$$ \rho_R(k_0) = { q^n \over 1-q^n } = { e^{- k_0 \over 2 T_R } \over
1 - e^{ - k_0 \over 2 T_R } }~.
$$
We can read off the ``right moving'' temperature
$$ T_R = {1 \over \pi}   {1 \over R }
 \sqrt{ \delta N_R \over Q_1 Q_5 }~.
$$
Now using \dbranes\ we find 
\eqn\hawkingstring{
T_R = { r_0^2 R_9 V \over2 g S_e }
}
in the near-extremal case, when $r_0$ very small. 
There is a similar factor for the left movers $\rho_L$
with a similar looking temperature
\eqn\tleft{ T_L = {1  \over \pi}   {1 \over R }
 \sqrt{  N \over Q_1 Q_5 }~.
}
The two temperatures $T_{L,R}$ can be thought of as the
effective temperature of the gas of left movers and the
gas of right movers. They are different because the gas
carries some net momentum. 
Since  $T_R \ll T_L$ the typical energy of the
outgoing string will be $k_0 \sim T_R $ and
$k_0/ T_L \sim T_R/T_L \ll 1 $ so that
we could approximate
\eqn\rholeft{
 \rho_L \sim { 2 T_L \over k_0  }=
{ 2  \over \pi k_0 R }
\sqrt{ N \over Q_1 Q_5}~.
}
The expression for the rate then is, up to a numerical constant,
\eqn\ratenew{
d\Gamma   \sim  { d^4 k \over k_0 } { 1 \over p_0^R p_0^L R V }
|{ \cal A}|^2  Q_1 Q_5 R \rho_R(k_0)  \rho_L(k_0)
}
where ${\cal A}$ is the disc diagram result.
 The factor
 $Q_1 Q_5 R$ is a volume factor, which 
arises from the delta funtion of momenta in \rateunpol\
$ \sum_n \delta (k_0 - 2 n/RQ_1Q_5 ) \sim R Q_1 Q_5 $.
The final expression for the rate is then, using
\stringamplitude ,~
\rholeft ~in \ratenew , up to a numerical constant,
\eqn\ratefinal{ d\Gamma   \sim
{ g^2 \over R V }  \sqrt{ Q_1 Q_5 N }
  { e^{- k_0 \over 2 T_R } \over
1 - e^{ -k_0 \over 2 T_R } } d^4 k
\sim 
({\rm Area} )  { e^{- k_0 \over 2 T_R } \over
1 - e^{ -k_0 \over 2 T_R } } d^4 k
}
Note that the powers of $k_0$ have cancelled. 
We conclude that the emission is thermal, with a physical
Hawking temperature
\eqn\hawkingstring{
 T_H = 2 T_R 
}
which exactly matches the classical result \thwk  .
It is an interesting  result that the area appeared correctly in
\ratefinal . Actually,   the coupling constant coming from
the string amplitude $\cal A$   is hidden in the
expression for the area (area = $4 G_N^5 S $).
Of course, it will be very interesting
to calculate the   coefficient  in \ratefinal ~to
see whether it exactly matches the absorbtion coefficient
 of a large
classical black hole. However it is easy to see that the 
absorbtion coefficient is proportional to the area for 
energies higher than the inverse of the Schwarschild radius
where one can calculate the cross section just from the
behaviour of classical geodesics. The coefficient in 
\ratefinal\ involves the cross section at energies
much lower than the inverse of the Schwarschild radius, which
is basically related to the temperature of the left movers. 
This means that calculating 
the absorbtion coefficient for this energy
requires solving the Klein Gordon equation on the
black hole background.  
In fact the result should not
depend on the internal polarizations of the outgoing
particles\foot{ Recently this question was addressed in \wadia
\dasmathurone\ \dasmathurtwo  . 
They calculated \dasmathurone ~the precise numerical 
coefficient in \ratefinal\ ~
and they found that the two calculations agree when $R$ is very large. }
The string theory calculation for (1,5) strings could  be done
using techniques similar to those in \luma .
Note that  the gas of antibranes is also at 
the temperature \hawkingstring\ and they also seem to   contribute 
to  Hawking radiation.
Notice that if we were emitting a spacetime fermion then
the left moving string could be a boson and the right moving
string a fermion, this produces the correct thermal factor
for a spacetime fermion. The opposite possibility gives
a much lower rate, since we do not have the enhancement
due to the large $\rho_L$ \rholeft . Notice also that
when separation from extremality is very small, then the number
of right movers is small and the statistical arguments used
to derive \ratefinal\ fail. Classically this should happen when
the temperature is so low that the emission of one quantum at temperature
$T$ causes the temperature to change by a large amount. This
occurs when the specific heat is of order one corresponding to
a mass difference from extremality \limitations\
\eqn\gapclass{
\delta M_{min} \sim { {G_N^5} \over  r_e^4 }
}
for a \RN\ black hole, with  $r_e$ as in \balance\  . 
The D-brane approach suggests the
existence of a mass gap of order
\eqn\gapdbrane{
\delta M_{min}  \sim 
{1\over  Q_1 Q_5 R  }
}
which using \constants \balance\  scales like \gapclass .
This is an extremely  
 small energy for a macroscopic extremal black hole. In
fact,
 it is of the order of the kinetic energy that the black hole would
have, due to the uncertainty principle, if we want to measure its
position with an accuracy of the order of its typical gravitational
radius
 $r_e$: $\delta M \sim (\Delta p )^2/M$ with $\Delta p \sim 1/r_e $.

We now examine the range of validity of these approximations.
For the purposes of this argument, we take the compactification
radii to be of order $\alpha'$ and set  $\alpha'=1$. In this
case \balance ~implies $Q_1\sim Q_5\equiv Q$ and $Q \sim g N$.
Then, by \entropyext, we find that the area of the horizon is
$A\sim(g^2 N)^{3/2}$. In order for the classical black hole interpretation to
hold, this area has to be much larger than $\alpha'$, so $g^2 N \gg 1$.
Since we want  $g$ to be  small,  $N$ is very large. This seems to
invalidate the perturbative D-brane picture since
open string loop corrections are of order $g Q = g^2 N $, and,
due to the large number $Q$  of D-branes, they are likely to be large
\sv , \hstro , \polchinskinotes\ . We will try to argue that
open string corrections might in fact be suppressed.
We note that the loop will
be in a  nontrivial  background of open strings. In fact, this
background was crucial to obtain a small
five dimensional coupling constant and
 non-zero area, which implies that somehow the D-branes might be
``separated'' from each other.
We suspect that this background of open strings suppresses
open string loops, enabling us to get results off extremality.
This is, of course, something to be checked in detail.
It is clear, however, that there are some circumstances where
open string backgrounds suppress loop contributions. For
example, compare loop contributions of $n$ D-branes on top of
each other  and $n$ widely separated D-branes. The difference
is just a background of open string translational zero modes.
It could be that the modes related to the entropy and Hawking
radiation are weakly interacting while higher energy modes
might indeed recieve large corrections. 
In this
respect we might recall an entirely different, but apparently
analogous,  physical situation: electrons and nuclei in a metal. The low
energy thermodynamics can be fairly well
reproduced by considering the electrons to be free, though there
are lots of charges present. In that case we have a better understanding as
to which  physical questions can be answered by regarding the electrons as
free and which require taking into account the interactions.
Hopefully, further studies of these models will provide a similar
understanding for the black hole case.

Finally, the fact that the perturbative D-brane treatment of non-extremal
physics gives the right results strongly suggests that there is more
than a grain of truth here. We think it quite possible that open string
quantum corrections are not as large as suggested by naive estimates.
The skeptic is  entitled to disagree!


\subsec{Extremal and Near-Extremal  Four Dimensional Black Holes.}

The extremal black hole solution of type IIA compactified
on $T^6 = T^4 \times S_1' \times S_1 $  is constructed by
wrapping $Q_6$ D-sixbranes on the whole $T^6$, $Q_2$
 D-twobranes on $S_1' \times S_1 $, $Q_5$ 
solitonic fivebranes along $ T^4 \times S_1$ and 
momentum $N/R$ flowing along $S_1$. All charges $ Q_2,Q_5,Q_6,N $
 are integers \ms .
 The supergravity 
solution was given in \tenfourd . 
Note that if we momentarily set $Q_5 =0$
we just simply have a configuration which is
T-dual to the one we had for the five dimensional 
black hole, so that the entropy is 
$\sqrt{ Q_2 Q_6 N }$ and is calculated as before in 
terms of open strings going between the D-twobranes
and the D-sixbranes. Alternatively, as in \sv\ 
we could view the D-twobranes as instantons on the
D-sixbrane and then the entropy comes from 
putting some left moving momentum along one
direction in the 2+1 dimensional field theory 
whose target space is the moduli space of these
instantons $ ( T^4)^{Q_2 Q_6}/S(Q_2 Q_6) $ where
$S(q)$ is the permutation group of $q$ elements
\vafacount .

\vskip 1cm
\vbox{
{\centerline{\epsfysize=2.7in \epsfbox{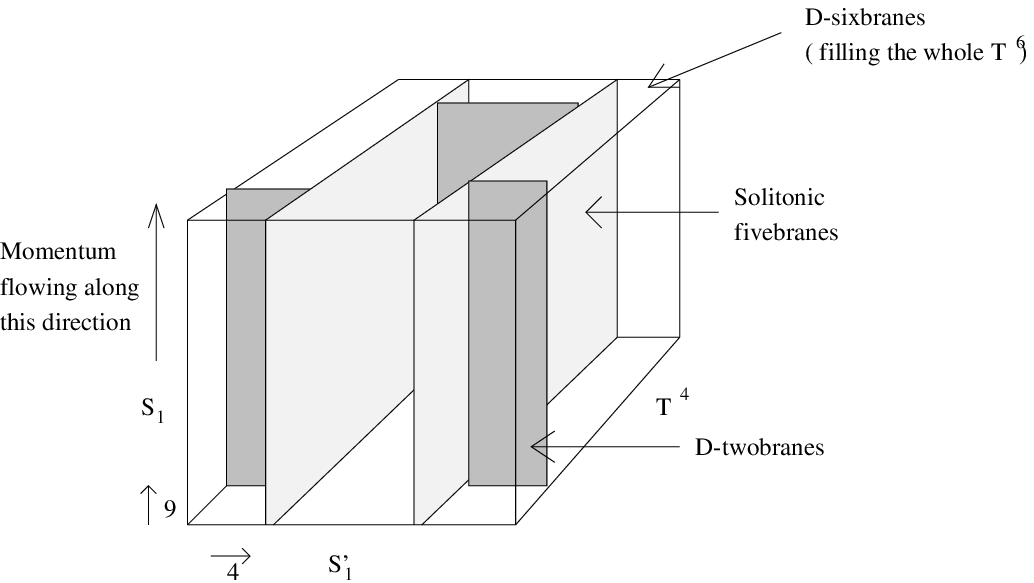}}}
{\centerline{\tenrm
FIGURE 8: Brane picture  of a  four dimensional black hole showing}}
{\centerline{ how the branes are wrapped in the compact dimensions.
}}
}
\vskip .5cm

Let us see what happens when we introduce the
fivebranes. The fivebranes intersect the 
two branes along the $S_1$. Different fivebranes
will be at different positions along the $S_1'$. 
The two branes can break and the ends separate in 
$T^4$ when it crosses the fivebrane. This was 
derived in 
\asop ,\twn\  and it follows, using
U-duality, from the fact that fundamental strings can end
on D-branes. 
Hence the $Q_2$ toroidal twobranes break
up into $Q_5 Q_2$
cylindrical twobranes, each of which is bounded by a pair of fivebranes. The
momentum-carrying open strings now carry an extra label describing which 
pair
 of
fivebranes
they lie in between. The number of species becomes $N_B=N_F=4Q_2Q_5Q_6$.
 The number of BPS-saturated states
of this system as a function of $Q_2,Q_5 ,Q_6$ and $N$ 
follows from the standard
$(1+1)$-dimensional entropy formula for a gas of massless particles
\eqn\std{S= 2 \pi \sqrt{\ (2 N_B+N_F  )E  R_9 \over 12 },}
where $N_B~~(N_F)$ 
is the number of species of right-moving bosons (fermions),
$E$ is the total energy and $2 \pi R_9$ 
is the size of the box. Using $ N_B = N_F =  4 Q_2 Q_5 Q_6 $
and $E=N/R_9$, we find the $R_9$-independent result for the large $N$
thermodynamic
limit 
\eqn\setn{S_{stat}=2\pi\sqrt{Q_2 Q_5 Q_6 N }.}
which indeed reproduces the classical result \entrofe .
This formula can be  justified using the usual BPS arguments.

This result can also be extended to black hole solutions
in $N=4$ supergravity by replacing $T^4$ in the previous
argument by $K3$ \sm . The entropy is the same as for the
$N=8$ case \setn .

Again, as in the five dimensional case, the calculation
leading to \setn\ implicitly assumed that the branes were 
singly wound  and it is  valid when $N$ is very large,
$N \gg Q_2 Q_5 Q_6$. 
For not so big values of $N$ the entropy comes from 
configurations where the branes are multiply wound.
In a fashion analogous to the five dimensional case this
leads to just 4 species of fermions and bosons propagating
on a circle of radius $ R Q_2 Q_5 Q_6 $ with the same result
\setn\ for the entropy.
Indeed, the classical energy gap for an extremal black hole
is \limitations\ 
\eqn\gapfour{ \delta M \sim { G_N^4 \over r_e^3 } \sim { 1 
\over R Q_2 Q_5 Q_6 }
}
which agrees with the multiply wound D-brane result. 

Now we consider near-extremal four dimensional black holes \hlm .
The simplest case to consider is when 
the size $R_9$ is much bigger than the rest of the
compact dimensions. This corresponds to taking
$ N_{\bar 2} \sim  N_{\bar 5} \sim
N_{\bar 6} \sim 0 $. Note that these antibrane
excitations are very massive when
$R_9$ is large.
When we have both left and right movers, using the
above arguments we find the entropy
\eqn\nearlong{
S = 2 \pi \sqrt{ N_2 N_5 N_6 } ( \sqrt{N_R} +  \sqrt{N_L} )~
}
which is the classical result \entropyb\ in the limit
we are considering.

Now we consider the case with angular momentum, again
in the limit of large $R_9$. 
With just twobranes and sixbranes present,
the D-brane excitations of this system are described
by a 1+1-dimensional field theory which turns out to be a
$(4,4)$ superconformal sigma model \spn .
The fivebrane breaks the right-moving supersymmetry
\harstro ,
leaving us with $(0,4)$ superconformal symmetry.
The $N=4$ superconformal algebra gives rise to
a left-moving  SU(2)$_L$ symmetry. Since fermionic
states in the sigma model become spinors in spacetime, the action
of O(3) spatial rotations has a natural
action on this SU(2)$_L$. The charge $F_L$ under
one  U(1) subgroup of this SU(2)$_L$  will then be
related to the four-dimensional angular
momentum (along one of the three axes)
 carried by the left movers by $J= F_L /2 $.
Due to the presence of the fivebrane the right-moving SU(2)$_R$
symmetry of the original $(4,4)$ superconformal field theory is broken and
the right movers cannot carry macroscopic angular
momentum.
The number of states with fixed
$N_L$, $N_R$, $F_L\gg 1$ may be computed as in
\spn ,\vbd\ to yield the entropy
\eqn\dent{
S= 2 \pi \sqrt{{ c\over  6 }} ( \sqrt{N_R} + \sqrt{\tilde N_L})
{}~,
}
where $\tilde n_L = n_L-
6 J^2/ c$ is the effective number of left movers that
one is free to change once one has demanded that
we have a given macroscopic angular momentum.
For our problem the central
charge is $c=6 N_2 N_5 N_6$ \ms, thus the
entropy \neent\ agrees with the D-brane formula
\dent .

It is interesting to take the extremal limit of these rotating
black holes, when the mass takes the minimum value consistent with
given angular momentum and charges. This happens when
$\tilde n_L =0 $, so the left movers are constrained to just carry the 
angular
momentum and do not contribute
to the entropy.
When the angular momentum is nonzero, even the extremal black hole is not
supersymmetric.
Using \dent\ and writing the result in
terms of the charge $N = N_R - N_L $ we find
\eqn\ehawken{
S=2\pi \sqrt{J^2 + N Q_2 Q_5 Q_6}~,
}
which indeed agrees with the entropy of an extremal
charged rotating black hole \cveticang .
Notice the surprising fact that although  we derived this formula
in the large $R_1$ regime (and  $J/M^2 \ll 1$), it continues to
be valid for arbitrary values of the parameters.
Since this is far from the BPS state, we had no reason to expect the
weak-coupling counting to continue to agree  with the black hole
entropy.

All the worries we had about possible strong coupling effects 
in five dimensions
are also a source of concern for these four dimensional black 
holes, but the successful calculation for the entropies 
encourages us to take this picture more seriously.
The discussion on Hawking radiation also carries 
over with almost no modification, giving the correct 
Hawking temperature.

\vfill\eject


\newsec{\bf DISCUSSION}

We have studied black holes in type II supergravity 
compactified to five and four dimensions. 
These theories have  several U(1) charges and 
 we considered
black holes that carry several of these charges. 
The extremal black holes have a direct correspondence with 
a superposition of string solitons and  they preserve some of
the supersymmetries.
For extremal black holes the entropy does not depend
on any of the continuous parameters and it is given
just in terms of the integer values of the quantized  charges.
In addition we constructed the non-extremal versions of
these black holes. This construction can be  done by 
applying U-duality symmetries and boosts to
the standard \RN\ solution in four or five dimensions.
We have shown how U-duality relates 
 the different  quantization 
conditions on the charges.
Black holes have some scalar charges which are determined,
due to  the no hair theorem, by the U(1) charges and the mass. 
If we compare the U(1) charges,
 the scalar charges,
and the mass with the corresponding values for a 
non-interacting collection of branes and antibranes we can
calculate the number of hypothetical 
 ``non-interacting'' constituents
of the black hole. 
The entropy formula has a  very suggestive
form when it is written 
in terms of these numbers. 

We then viewed the same collection  of branes and strings
from the point of view of string theory. 
Applying the rules for quantizing D-brane solitons we
were able to count the number of microscopic states of 
such configurations. In the extremal case this counting
can be justified 
rigorously by using the standard BPS arguments.
In the near-extremal case one can certainly do the
counting on the string theory side when the coupling is very
weak. However, in order to compare to the black hole answer
we need to make the coupling bigger. 
It turns out that the necesary size of the coupling is
such that one might expect  large corrections.
 However, the near 
extremal entropy is precisely accounted for by
this weak coupling argument. D-branes account for some
non-perturbative effects, so the question is whether they 
account for all the necessary ones to describe 
black holes. The answer is that they seem to be doing that,
at least as far as entropy calculations is concerned\foot{
See also \wadia \dasmathurone \dasmathurtwo\ ~regarding the
comparison of  scattering
amplitudes between the classical approach and the D-brane approach.}.
These near-extremal black holes
have many free paramenters, in fact one can also consider
solutions with angular momentum. For all  these cases it is
possible to account for the entropy, indicating that 
the understanding of the  black hole degrees of freedom
seems to be not too far from reality.
The energy gap for excitations of an extremal  black hole also 
agrees using both methods (classical and D-brane). 

Hawking radiation is viewed in this approach as
the collision of two oppositely moving open strings
attached to the branes that decay into a closed string that
leaves the brane. As expected from thermodynamics arguments,
the Hawking temperature is precisely the classical 
result. Furthermore, the Hawking radiation rate shows 
that the D-brane calculation ``knows'' about the geometry, since
the rate is proportional to the area of the black hole. 
The overall coefficient in this rate is proportional to 
the absorbtion cross section of the  black hole for that particular
mode. 
It will be interesting to study 
these more dynamical questions to understand better whether this
object really represents a black hole or not.

One might worry that we are not  considering a 
realistic conpactification since our world is not
 described by $N=8$ supergravity, at least 
at small energies. It will be indeed very interesting 
to extend these results for the case of $N=2,1,0$ and see
how much of this description carries over.
From a purely theoretical point of view, the 
problem of black hole entropy is  as puzzling in 
$N=8$ supergravity as it is in General Relativity, 
but it is easier to analyse in $N=8$ supergravity.
All our results carry over to the $N=4$ theory, both in
their type II version (type II on K3) as in the 
heterotic theory (type I) compactified on a torus. 

On a more speculative note  we would say that 
if this picture were qualitatively right, there would be no 
information loss, the information would stay  on the open strings
that live on the branes which sit at the horizon (for an extremal 
black hole). In this
picture, the classical region inside the horizon would be
an effective description of the dynamics of these open strings.
What would happen is that an observer, made of closed 
strings,   that 
falls through the horizon  is turned into open strings 
together with all his measuring apparatus, so it seems plausible
that he would not notice the difference. There should be 
a way to describe the subsequent dynamics in terms of
some effective closed strings that falled through the horizon. 
 Note that
the problem of information loss could be analysed in 
terms of near-extremal black holes. 
Because of our lack of control on the strong coupling
problem we cannot say anything definite about information
loss.
This D-brane description is the description of the physics as 
seen by the asymptotic  observer. It is for this observer that
evolution should be unitary since he sees the black hole formation
and evaporation process.

Let us end by saying that black holes are an excellent theoretical 
laboratory
for understanding some features of quantum gravity. One could say that
they are the ``Hydrogen atom'' of quantum gravity. 
It will be interesting to see what  string theory  will say about
this in the future.

\vfill\eject

\listrefs

\bye